\documentclass[abstract]{aa}
\usepackage[T1]{fontenc}
\usepackage{ae,aecompl}
\usepackage{graphicx}
\usepackage{txfonts}
\usepackage{amsmath}
\usepackage{amssymb}
\usepackage{sidecap}
\usepackage{xcolor}
\usepackage{subfig}
\usepackage{appendix}
\usepackage{rotating}
\bibpunct{(}{)}{;}{a}{}{,} 
\def \msun{\rm \, M_\odot}
\usepackage{multirow}

\titlerunning{Non-thermal filaments and AGN recurrent activity in the galaxy group Nest200047} 
\authorrunning{M. Brienza, et al.}

\usepackage{hyperref}
\usepackage{upgreek}

\hypersetup{
  colorlinks   = true, 
  urlcolor     = blue, 
  linkcolor    = red, 
  citecolor    = blue 
}

\begin{document} 


\title{Non-thermal filaments and AGN recurrent activity in the galaxy group Nest200047: a LOFAR, uGMRT, MeerKAT, VLA \\ radio spectral analysis}

\author{M. Brienza$^{1,2}$, K. Rajpurohit$^{3}$, E. Churazov$^{4,5}$, I. Heywood$^{6,7,8}$, M. Br\"uggen$^{9}$, M. Hoeft$^{10}$, F. Vazza$^{2,1}$, A. Bonafede$^{2,1}$, A. Botteon$^{1}$, G. Brunetti$^{1}$, F. Gastaldello$^{11}$, I. Khabibullin$^{12,4,5}$, N. Lyskova$^{5,17}$, A. Majumder$^{13}$, H.J.A. R\"ottgering$^{14}$, T.~W.~Shimwell$^{15,14}$, A. Simionescu$^{13,14,16}$, R. J. van Weeren$^{14}$}

\bigskip

\institute{Istituto Nazionale di Astrofisica (INAF) - Istituto di Radioastronomia (IRA), via Gobetti 101, 40129, Bologna, Italy
\and
Dipartimento di Fisica e Astronomia, Università di Bologna, via P. Gobetti 93/2, I-40129, Bologna, Italy
\and
Center for Astrophysics-Harvard and Smithsonian, 60 Garden Street, Cambridge, MA 02138, USA
\and
Max Planck Institute for Astrophysics, Karl-Schwarzschild-Str. 1, Garching b. M\"unchen 85741, Germany
\and
Space Research Institute (IKI), Russian Academy of Sciences, Profsoyuznaya 84/32, 117997 Moscow, Russia
\and 
Astrophysics, Department of Physics, University of Oxford, Keble Road, Oxford, OX1 3RH, UK
\and
Centre for Radio Astronomy Techniques and Technologies, Department of Physics and Electronics, Rhodes University, PO Box 94, Makhanda 6140, South Africa
\and
South African Radio Astronomy Observatory, 2 Fir Street, Black River Park, Observatory 7925, South Africa
\and
Hamburger Sternwarte, University of Hamburg, Gojenbergsweg 112, 21029, Hamburg, Germany
\and
Thüringer Landessternwarte, Sternwarte 5, 07778 Tautenburg, Germany
\and 
IASF – Milano, INAF, Via A. Corti 12, I-20133 Milano, Italy
\and
Universit\"ats-Sternwarte, Fakult\"at f\"ur  Physik, Ludwig-Maximilians Universität, Scheinerstr. 1, 81679 M\"unchen, Germany
\and
SRON Netherlands Institute for Space Research, Niels Bohrweg 4, 2333CA, Leiden, The Netherlands
\and
Leiden Observatory, Leiden University, PO Box 9513, 2300 RA Leiden, The Netherlands
\and
ASTRON, Netherlands Institute for Radio Astronomy, Oude Hoogeveensedijk 4, 7991 PD, Dwingeloo, The Netherlands
\and
Kavli Institute for the Physics and Mathematics of the Universe (WPI), The University of Tokyo, Kashiwa, Chiba 277-8583, Japan
\and 
Astro Space Center, Lebedev Physical Institute, Russian Academy of Sciences, Russia}

\date{Accepted ---; received ---; in original form \today}

\abstract{Nest200047 is one of the clearest examples of multiple radio bubbles from an Active Galactic Nucleus (AGN) observed in a galaxy group, also featuring a complex system of non-thermal filaments, likely shaped by buoyancy and gas motions in the group and stabilized by large-scale magnetic fields.

In this study, we present a new set of high-quality data obtained from our dedicated observational campaign using uGMRT, MeerKAT, and VLA. By combining this with existing LOFAR data, we perform a detailed morphological and spectral analysis of the system over a broad frequency range (53-1518 MHz) using various complementary techniques.

Our images reveal new filamentary emission in the inner 60 kpc of the system, surrounding and extending from the inner bubbles and jets, suggesting a complex dynamical evolution of the non-thermal plasma in the group core too. Overall, all filaments in the group have a width of a few kpc and lengths from a few tens up to a few hundred of kpc. They all show a steep ($\rm \alpha=1\sim2$) and curved (spectral curvature up to 1) radio spectrum.  Interestingly, the filaments exhibit a constant radio spectral index profile along their length, indicating that particles are not cooling along them. This suggests that the particles were either (re-)accelerated together and evolved in a similar magnetic field or they are moving along the filaments at super-Alfvenic speeds.
A spectral ageing analysis based solely on radiative losses provides age estimates for the three different pairs of bubbles of 130 Myr, 160–170 Myr, and >220 Myr, with jet active times ranging between 50 and 100 Myr, and very short inactive times. This supports the idea of a nearly continuous energy injection, typical of the "maintenance mode" of AGN feedback, particularly in galaxy groups. By taking into account adiabatic expansion, we also find that the outermost northern bubble cannot be considered a simple evolution of the inner bubbles unless some re-acceleration or mixing with the external gas is invoked or, alternatively, different outburst properties.

In conclusion, our study clearly shows the potential of the combined use of high-quality, new-generation radio data for understanding recurrent jet activity and feedback, and anticipates the new opportunities that will be offered by the SKA observatory.

\noindent
}

\maketitle

\section{Introduction}
\label{intro}

Jets from Active Galactic Nuclei (AGN) recurrently inflate lobes of cosmic rays and magnetic fields extending on various scales and up to a few Mpc. Over the past few decades, it has become evident that the energy released by these jets can significantly influence both the properties of the host galaxy and the thermodynamic evolution of the intragroup medium (IGrM) or intracluster medium (ICM) in which the galaxy resides. In particular, this so-called jet-induced feedback is invoked to prevent the overcooling of the IGrM/ICM (e.g., \citealt{churazov2000}, see reviews by \citealt{fabian2012,mcnamara2012,hlavaceklarrondo2022}) and, in turn, to regulate star formation of the massive galaxies ($\rm>10^{11} \msun$) that lie at the centre of galaxy groups and clusters \citep[e.g.][]{croton2006}. 

However, important aspects, such as what is the duty cycle of the jet activity in different environments and host galaxies, and how the released energy gets distributed and eventually thermalized in the surrounding medium, are still under debate. The most popular mechanisms proposed for energy transfer are shocks \citep[e.g.][]{fabian2003,li2017}, sound waves \citep[e.g.][]{fabian2005} internal waves \citep[e.g.][]{zhang2018}, turbulence \citep[e.g.][]{zhuravleva2014}, streaming of cosmic rays \citep[e.g.][]{ehlert2018, ruszkowski2023}, and mixing \citep{hillel2017}. Even if the majority of the energy is released in the form of outbursts, simulations show that in the long run, AGN lobes are still able to heat the ICM and gently raise the core entropy \citep[e.g.][]{bourne2021}. How these lobes evolve on long timescales in the IGrM/ICM after the jets have switched off and how they do eventually mix with it is also poorly understood though.

Theories show that, when located in a galaxy group or cluster, the old (so-called remnant) lobes are expected to rise buoyantly as light bubbles through the local atmosphere and be transformed into toroidal structures due to the encountered pressure gradient \citep[e.g.][]{churazov2000,bruggen2001}, resembling mushroom clouds produced by a powerful explosion in a stratified atmosphere. Depending on the properties of the system (such as AGN jet power and the IGrM/ICM pressure gradient) and on the plasma microphysics (including magnetic fields, cosmic rays, and viscosity), the rising torus may either get completely disrupted by Rayleigh-Taylor and Kelvin-Helmholtz instabilities, or rise to a height where the effective entropy of a mixture of non-thermal and thermal plasmas matches that of the ambient IGrM/ICM, subsequently spreading laterally. Factors such as high viscosity, magnetic fields, and cosmic ray protons within the bubble have been found to be essential to stabilize the bubbles, preventing their complete fragmentation over timescales of a few Myr \citep[e.g.][]{reynolds2005, deyoung2003,  ruszkowski2017, ehlert2018, yang2019} 

On top of this, the evolution of the remnant lobes can also be altered by ambient "cluster weather", i.e. the fluid motions in the ICM promoted by sloshing and/or mergers. This can displace the lobe content from its original position, promoting its disruption and the mixing of hot lobe material with the hot atmosphere, especially after the jet has switched off \citep[e.g.][]{mendygral2012, bourne2021, zuhone2021, fabian2021, vazza2021, brienza2022, botteon2024}. Eventually, the lobes content is expected to get spread and to provide a non-negligible reservoir of magnetic fields and non-thermal particles in the IGrM/ICM, with possible implications also for the formation of diffuse radio sources in merging galaxy clusters \citep{vanweeren2019}.

\begin{figure*}
\centering
\includegraphics[width=1\textwidth]{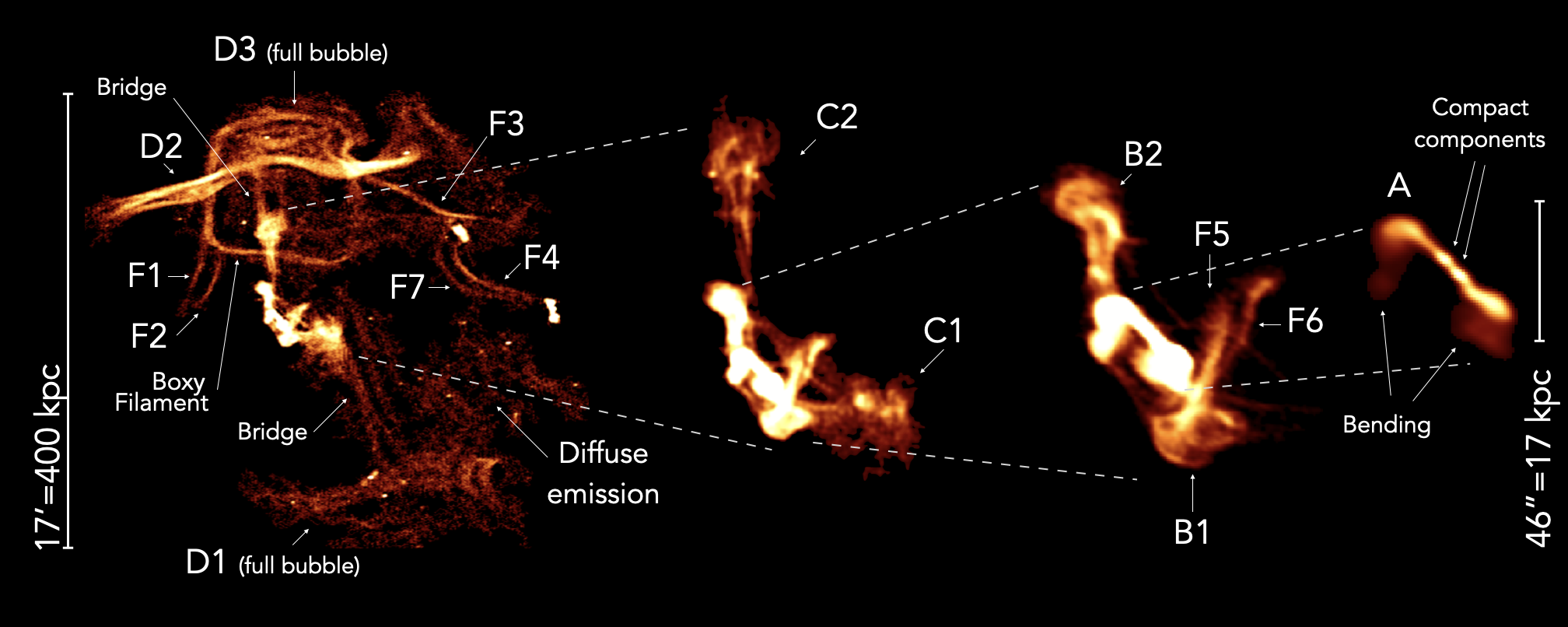}
\caption{Highlight of the main morphological radio structures on different scales observed in the galaxy group Nest200047, with labels. From the left to the right: LOFAR image at 144 MHz and 6-arcsec resolution, uGMRT image at 400 MHz with 6-arcsec resolution, and uGMRT image at 700 MHz with 5-arcsec resolution.}
\label{fig:presentation}
\end{figure*}

Direct and detailed observations of this evolutionary process have always been challenging. On the one hand, the lifetime of the relativistic particles producing the radio emission is often much shorter than the time required for the bubble to rise and transform in the IGrM/ICM. This is especially true for galaxy groups, where expansion losses are expected to be much more significant than in more massive systems. On the other hand, high spatial resolution and sensitivity is necessary, which limits observations to very nearby and/or intrinsically large systems. In the last years, however, the advent of a new generation of radio interferometers, particularly sensitive at low frequencies, has opened up a new window for the study of the oldest radio-emitting AGN bubbles.

In this context, the galaxy group Nest200047, originally identified in the Two Micron All Sky Survey (2MASS) by \cite{tully2015} at redshift z=0.01795, represents a unique testbed to study recurrent AGN activity and the physics and fate of old AGN bubbles in a galaxy group. Indeed, using Low Frequency Array (LOFAR, \citealp{vanhaarlem2013}) in the frequency range 50-150 MHz, \cite{brienza2021} first reported very complex radio emission associated with the system extending over more than 400 kpc (see Fig. \ref{fig:presentation}), which was interpreted as a series of bubbles ejected by the central black hole of the brightest group galaxy (BGG) MCG+05-10-007 (K-band magnitude equal to m$_{\rm k}$=8.87 and stellar mass equal to logM$\rm _{*}/\msun$=11.56). The bizarre boxy filament of the oldest and largest of these bubbles, with an estimated age of 300-400 Myr, recalls a much more evolved version of the “mushroom-shaped” bubble observed in M87 on kpc scales \citep[e.g.][]{owen2000, churazov2001, degasperin2012}. Based on observations performed with the extended ROentgen Survey with an Imaging Telescope Array (eROSITA, \citealp{predehl2021}) on board of the Spectrum-Roentgen-Gamma (SRG) mission \citep{sunyaev2021} the X-ray luminosity, temperature and mass of the system were estimated to be $\rm L_X=5-10\times10^{42} \ erg \ s^{-1}$, $\rm kT_X\sim$2~keV and $\rm M_{500}=3-7\times10^{13} \ M_{\odot}$, respectively \citep{brienza2021}.

Nest200047 clearly shows that although the old AGN-produced plasma is stretched over a large distance of $\sim$400 kpc, it remains poorly mixed with the ambient IGrM even, suggesting that complete thermalization of their energy has yet to occur. Instead, the old bubbles appear to be deformed into intricate filamentary structures, likely stabilized by magnetic fields. Since the discovery of Nest2000247, the detection of such filamentary structures in various astrophysical contexts has grown substantially but their properties are still poorly constrained and their nature is unclear, making their detailed investigation essential.

In this paper, we present a thorough radio broadband spectral analysis of the galaxy group Nest200047 aimed at understanding the physics of the system. For this, we combine the already-published LOFAR data with new datasets from our dedicated observational campaign conducted with the upgraded Giant Metrewave Radio Telescope (uGMRT), MeerKAT \citep{jonas2016}, and Karl G. Jansky Very Large Array (VLA) covering the frequency range 53-1500 MHz. A polarization analysis of the source will be presented in a subsequent work.

The outline of the paper is as follows: In Sect. \ref{sec:data} we describe the data used in the work, and the data calibration and imaging strategies. In Sect. \ref{sec:results} we present the radio images and detailed spectral analysis, including spectral index maps and profiles, colour-colour plots, tomography maps, global spectra, spectral ageing, and surface brightness correlations. A discussion and summary of our main findings are finally given in Sect. \ref{sec:discussion} and Sect. \ref{sec:concl}, respectively.

The cosmology adopted throughout the paper assumes a flat universe with the following parameters: $\rm H_{0} = 70$ $\rm km$ $\rm s^{-1}$ $\rm Mpc^{-1}$, $\rm \Omega_{\Lambda} =0.7$, $\rm \Omega_{M} =0.3$. At the redshift of MCG+05-10-007 ($z=0.01795$) the luminosity distance is 78.4\,Mpc and 1$\arcsec$ corresponds to 0.367\,kpc. All coordinates are given in J2000. The spectral index $\alpha$ is defined as $S \propto \nu^{-\alpha}$.

\section{Observations and data processing}
\label{sec:data}

We performed dedicated follow-up observations of Nest200047 using uGMRT, MeerKAT, and VLA. In this section, we describe the calibration and imaging strategies used for each dataset. For the analysis presented in this work, we also use the LOFAR data already presented in \cite{brienza2021}. A summary of all observations and images used is presented in Table \ref{tab:data} and Table \ref{tab:imageparam}, respectively. Images are shown in Fig. \ref{fig:high}, \ref{fig:zoom}, \ref{fig:mid} and \ref{fig:low}.

\begin{table*}[!htp]
\centering
        \small
 \caption{Summary of the observations of Nest200047 used in this work.
 }

                \begin{tabular}{c c c c c c c c c}
                \hline
                \hline
                Telescope & Bandwidth & Central frequency & UVmin & TOS$^1$ & Integration time & Date & Project Code\\
                & [MHz] & [MHz] & & [hours] & [seconds] & & &\\
                \hline
                \hline
                LOFAR$^{2}$ & 30-78  & 53  & 12$\lambda$ &8 & 1 & 17/04/2020 & DDT13\_003\\
                LOFAR$^{2,3}$ & 120-168  & 144  & 32$\lambda$ &16 & 1 & 27/03/2020, 19/08/2020 & LC12\_015, LT14\_004\\ 
                uGMRT & 300-500  & 400  & 100$\lambda$ &9.7 & 5.3 & 21/11/2020 & 39\_010\\
                uGMRT & 550-950  & 700  & 200$\lambda$ &9.7 & 5.3 & 19/11/2020 & 39\_010\\
                MeerKAT & 544-1087  & 816  & 50$\lambda$ &4.7 & 7 & 12/02/2022 & DDT-20211020-IH-01\\
                MeerKAT & 856-1711  & 1284 & 80$\lambda$ &4.4 & 7 & 25/01/2022 & DDT-20211020-IH-01\\ 
                VLA B array & 1000-2000  & 1518  & 700$\lambda$ &4.5 & 3 & 25/09/2021 & 21A-065\\
                VLA C array & 1000-2000  & 1518  & 250$\lambda$ &3.1 & 3 & 08/06/2020, 27/06/2021 & DDT20A-442/21A-065\\
                VLA D array & 1000-2000  & 1518  & 180$\lambda$ &2 & 3 & 02/04/2021 & 21A-065\\
                \hline
                \hline  
                \end{tabular}
                \tablefoot{$^1$Time On Source; $^2$Observations presented in \cite{brienza2021}; $^3$Observations not centered on the target.}
\label{tab:data}
\end{table*}

\begin{table*}[!htp]

\small
\caption{Summary of the radio images of Nest200047 presented in this work (see Fig. \ref{fig:high}, \ref{fig:zoom}, \ref{fig:mid} and \ref{fig:low}).}
\centering

\begin{tabular}{c c c c c}
\hline
\hline
Central frequency & Beam & Weighting & Outer UV-taper & RMS noise\\
$\rm[MHz]$ & [arcsec$\times$arcsec]& & [arcsec] & [$\mu$Jy/beam]\\
\hline
53  & 9.2$\times$14.6 & Briggs -0.8 & - & 1500 \\
53  & 14.5$\times$24.2 & Briggs -0.5 & 10 & 1500 \\ 
\hline
144  & 4.3$\times$8.6 & Briggs -0.5 & - & 166 \\
144  & 10$\times$10$^1$ & Briggs -0.5 & - & 230 \\
144  & 18.2$\times$20.2 & Briggs -0.5 & 15 & 260 \\
\hline
400  & 4.8$\times$5.8 & Briggs -0.5 & - & 30 \\
400  & 10.5$\times$11.5 & Briggs 0 & 8 & 45 \\
400  & 22.6$\times$23.8 & Briggs 0 & 20 & 100 \\
\hline
700  & 3.1$\times$5.0 & Briggs 0 & - & 9 \\
700  & 10$\times$10$^1$ & Briggs 0 & - & 27 \\
700  & 14.1$\times$24.9$^2$ & Briggs 0 & 25 & 60 \\
\hline
815  & 6$\times$13.6 & Briggs -0.8 & - & 25 \\
815  & 15$\times$15$^3$ & Briggs -0.8 & - & 35 \\
815  & 9.2$\times$21.6$^2$ & Briggs 0 & - & 50 \\
\hline
1280  & 4.1$\times$9.4 & Briggs -0.5 & - & 7 \\
1280  & 5.4$\times$12.4$^3$ & Briggs 0 & - & 8 \\
1280  & 20.7$\times$23.8 & Briggs 0 & 15 & 22 \\
\hline
1518  & 3.4$\times$4.3 & Briggs 0 & - & 8 \\
1518  & 12.8$\times$13.5 & Briggs 0 & 12 & 7 \\
1518  & 25.4$\times$25.7 & Briggs 0 & 25 & 20 \\
\hline
\hline  
\end{tabular}
\tablefoot{$^1$Smoothed to 10 arcsec; $^2$Smoothed to 25 arcsec; $^3$Smoothed to 15 arcsec.}
\label{tab:imageparam}
\end{table*}

\subsection{LOFAR 54 MHz and 144 MHz}
\label{lofar}

LOFAR observations of Nest200047 with both the Low Band Antennas (LBA, 30-78 MHz) and the High Band Antennas (HBA, 120-168 MHz) are presented in \cite{brienza2021}, to which we refer the reader for details on the calibration procedures used. For this work, the previously calibrated data were used to produce new images with parameters matching the datasets at higher frequencies as summarised in Table \ref{tab:imageparam} and with appropriate uv-cuts using {\tt WSClean} (version 2.8, \citealp{offringa2014}) as discussed in Sect. \ref{sec:spix}.

\subsection{uGMRT 400 MHz and 700 MHz}
\label{gmrt}

We observed Nest200047 with the upgraded Giant Metrewave Radio Telescope (uGMRT) during November 2020, in both band 3 (300-500 MHz) and band 4 (550-950 MHz). The total on-source observing time was 9.7 hours in each band. 3C48 and 3C147 were used as primary calibrators and observed for 8 minutes at the beginning and the end of each observing run, respectively. No phase calibrator was observed. At both frequencies, the total bandwidth was divided into 4096 channels and the integration step was set to 5.3 seconds.
We calibrated data and performed direction-dependent self-calibration using the {\tt SPAM} pipeline \citep{intema2014, intema2017} upgraded for handling new wideband uGMRT data\footnote{\url{http://www.intema.nl/doku.php?id=huibintemaspampipeline}} and we set the absolute flux scale according to \cite{scaife2012}. Due to severe radio frequency interference (RFI), data above 850 MHz in band 4 were discarded. Using the output calibrated data we created the final images with multiscale cleaning in {\tt WSClean}.

\subsection{Meerkat 815 and 1218 MHz}
\label{meerkat}

Nest200047 was observed in the UHF (544--1017~MHz; 4.7~h on-source) and L (856–-1711 MHz; 4.4~h on-source) bands with 4k channels using the MeerKAT telescope \citep{jonas2016} in January and February 2022. For both observations, the primary calibrator was PKS B0408-65, which we used to calibrate the instrumental delays, bandpass, and set the absolute flux scale according to \cite{reynolds1994}. We note that, as explained in \cite{riseley2022}, this is tied to the \cite{baars1977} flux scale and has an offset from the \cite{scaife2012} flux scale of only $\rm\sim3\%$. The secondary calibrator in both cases was J0403+2600, which was used to calibrate the time-dependent complex gains by observing it for 1.5 minutes for every 20 minute on-source scan. Both observations were processed using {\tt oxkat} \citep{heywood2020}, a set of semi-automated scripts for imaging MeerKAT data, running on the \emph{ilifu} cluster\footnote{\url{https://www.ilifu.ac.za/}}. The scripts employ several radio astronomy packages, including {\tt CASA} \citep{mcmullin2007} for reference calibration, {\tt tricolour} \citep{hugo2022} for flagging, {\tt WSclean} for imaging, {\tt cubical} \citep{kenyon2018} for self-calibration, and {\tt shadems} \citep{smirnov2022} for diagnostic plots. In particular, one round of phase-only direction-independent self-calibration was performed. The software packages are containerised using {\tt singularity} \citep{kurtzer2017}. 
A detailed description of the data processing is provided by \cite{heywood2022c}.

\subsection{VLA 1518 MHz}
\label{vla}

We observed Nest200047 with the VLA in B, C, and D configurations aiming at obtaining a spatial resolution comparable with the other datasets and at the same time the highest sensitivity possible to large-scale emission. The observations were performed June 2020 and in April, June, September 2021. The total time on source combining BCD arrays was $\sim$9.5 hours and 1-GHz bandwidth was recorded, spread over 16 spectral windows each composed of 64 channels. 3C138 was used as primary calibrators and J0414+3418 was used as a phase calibrator and observed regularly every 15 minutes. 
Each configuration dataset was first calibrated separately following the standard data reduction scheme in {\tt CASA}. This includes RFI flagging, bandpass calibration, delays calibration and gain calibration. Further RFI flagging was performed using {\tt AOFlagger} \citep{offringa2012}. The flux density scale was set according to \cite{perley2013}. The resulting calibrated data were averaged by a factor of 4 in frequency per spectral window and in time intervals of 10s, 10s, 6s in time for D, C and B configurations, respectively. We then performed a few rounds of phase self-calibration and a final round of amplitude-phase self-calibration on each dataset separately. Finally, all configurations data were imaged together using {\tt WSClean}, with multi-scale clean.

\section{Results}
\label{sec:results}

\subsection{Radio morphology}
\label{sec:morpho}

\begin{figure*}[!htp]
\centering
\includegraphics[width=1\textwidth]{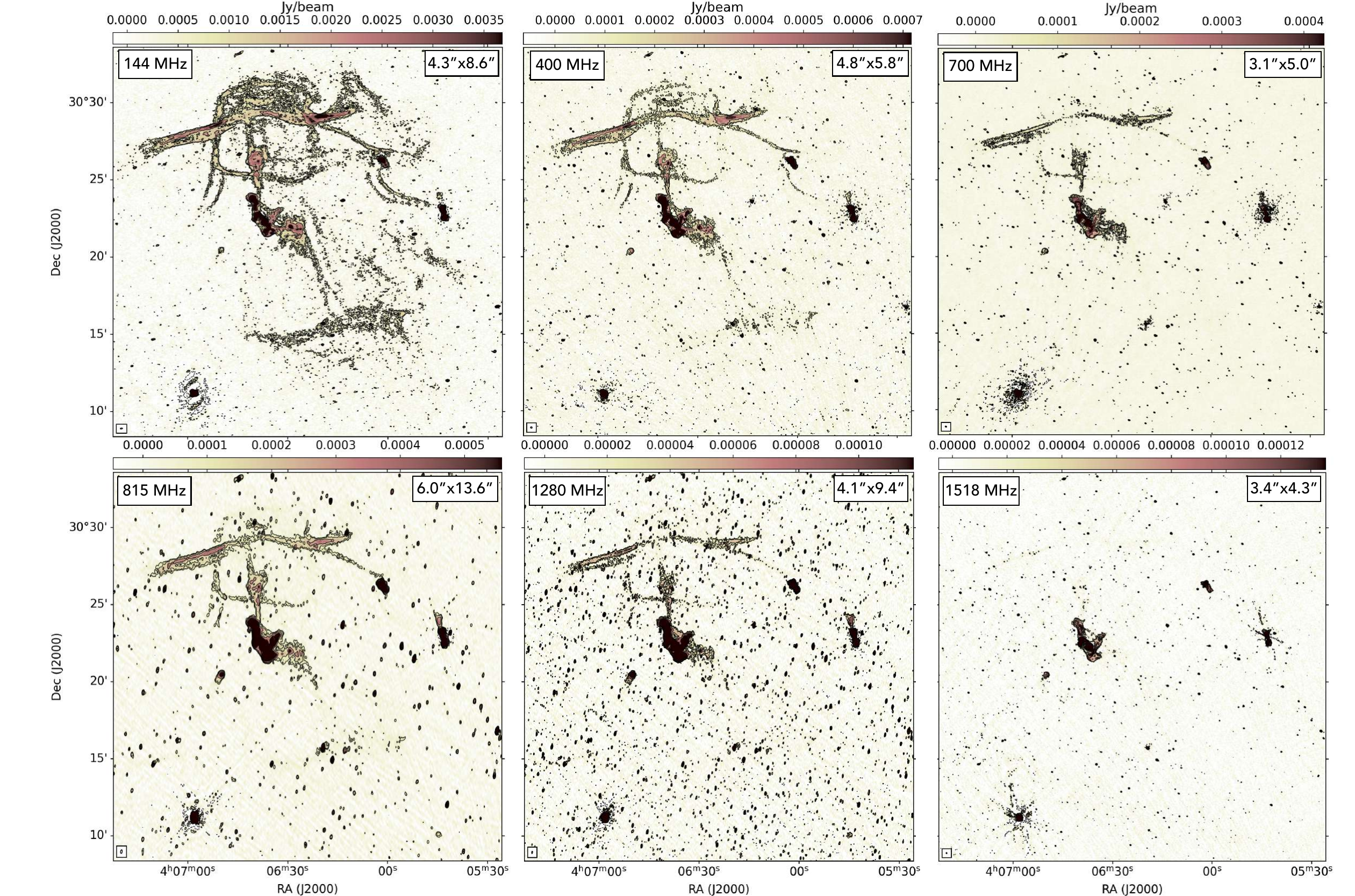}
\caption{Set of images at the highest angular resolution with increasing frequency from left to right, in the range 144-1518 MHz (see Table \ref{tab:imageparam} for details on the imaging parameters and noise values). Contours are drawn at -3, 3, 5, 10, 20 $\rm \times \ \sigma_{local}$. Labels for the main morphological features are shown in top-left panel. }
\label{fig:high}
\end{figure*}

\begin{figure*}[!htp]
\centering
\includegraphics[width=0.9\textwidth]{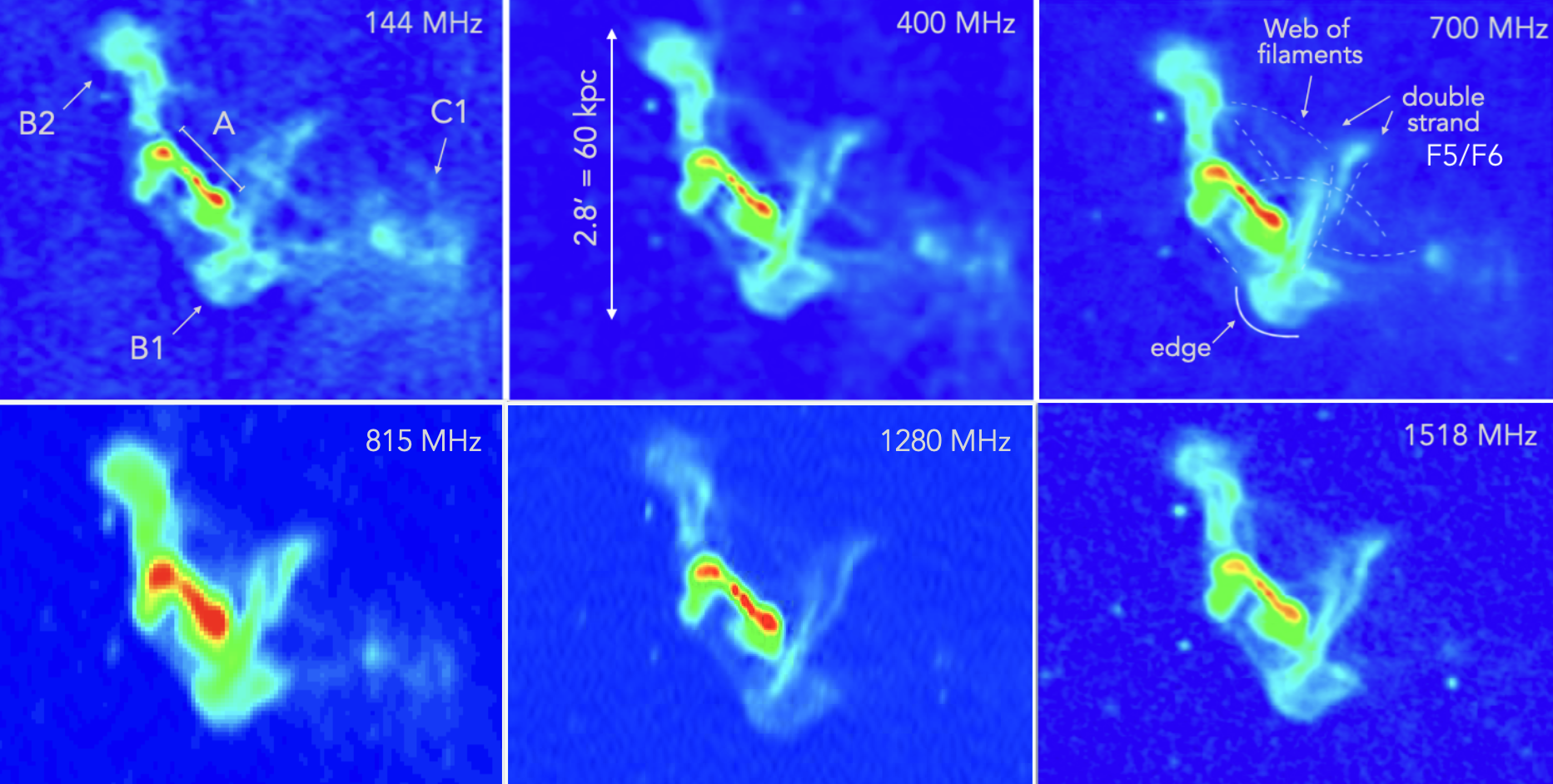}
\caption{Zoom-in on the central region in the images at the highest available resolution, namely 144 MHz, 400 MHz, 700 MHz and 1518 MHz. Labels are added for the main morphological features. All panels are zoomed into the same scale.}
\label{fig:zoom}
\end{figure*}

\begin{figure*}
\centering
\includegraphics[width=1\textwidth]{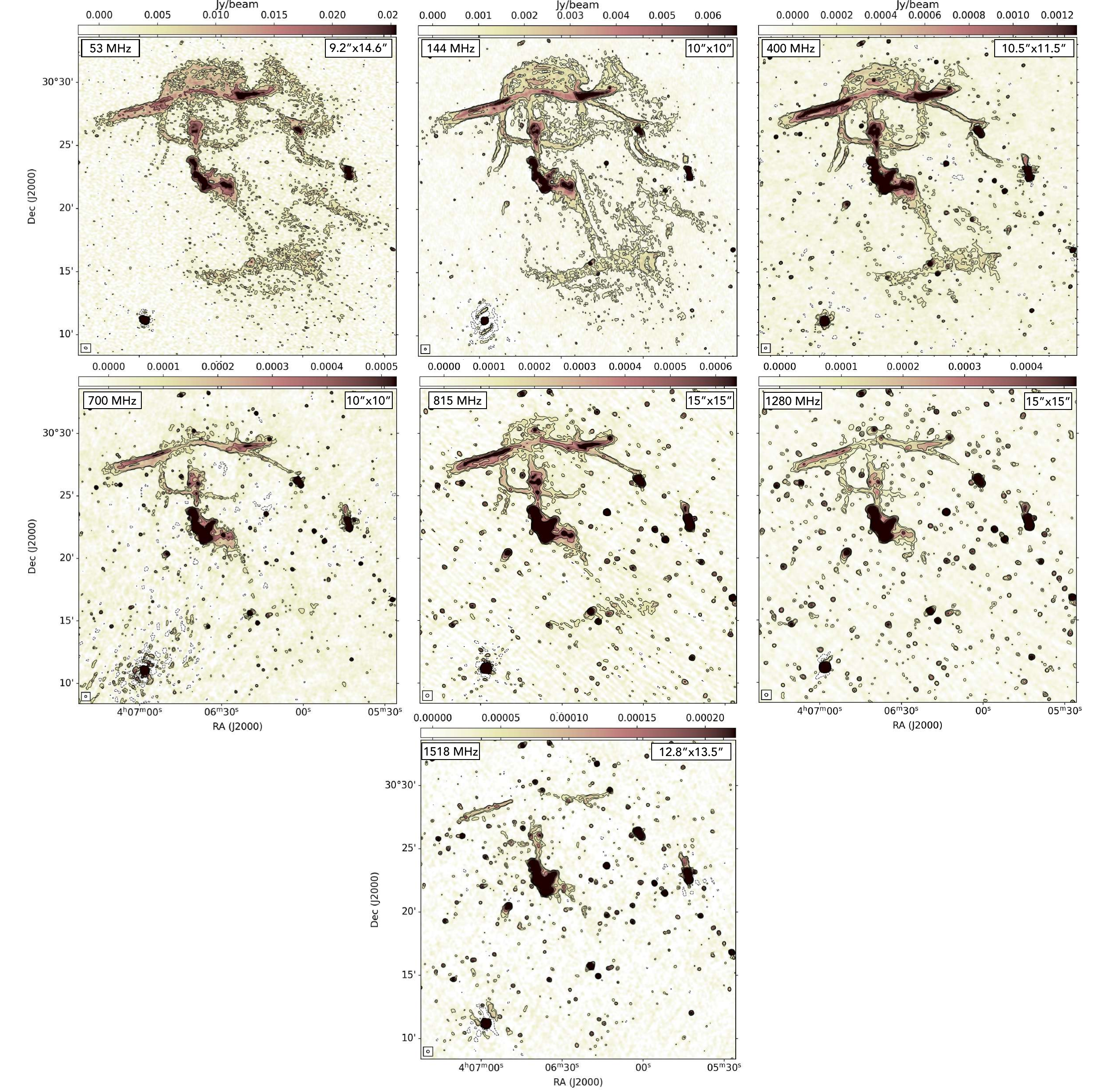}
\caption{Set of images at intermediate angular resolution with increasing frequency from left to right, in the range 53-1518 MHz (see Table \ref{tab:imageparam} for details on the imaging parameters and noise values). Contours are drawn at -3, 3, 5, 10, 20 $\rm \times \ \sigma_{local}$.}
\label{fig:mid}
\end{figure*}

\begin{figure*}
\centering
\includegraphics[width=1\textwidth]{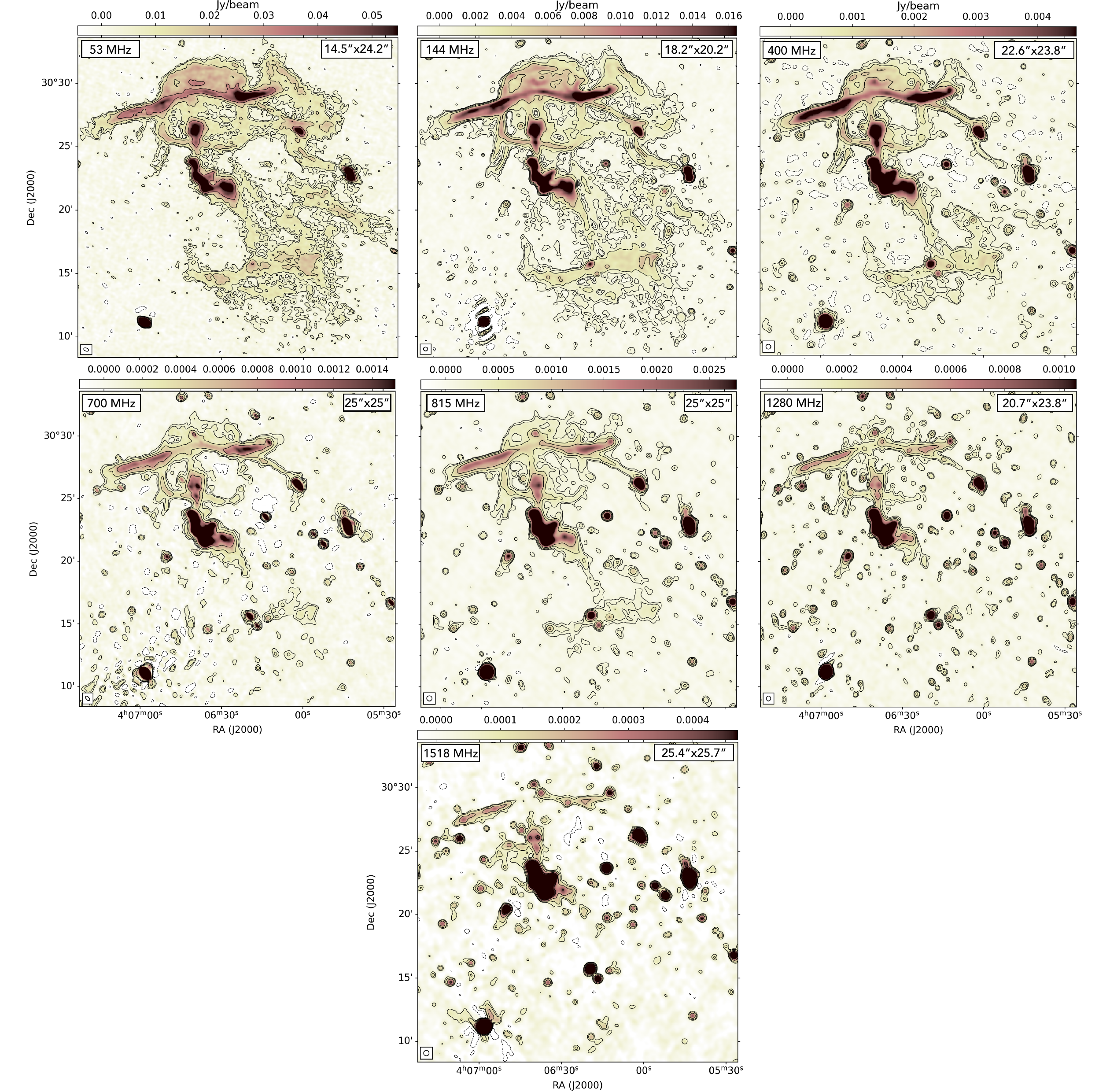}
\caption{Set of images at low angular resolution with increasing frequency from left to right, in the range 53-1518 MHz (see Table \ref{tab:imageparam} for details on the imaging parameters and noise values). Contours are drawn at -3, 3, 5, 10, 20 $\rm \times \ \sigma_{local}$.}
\label{fig:low}
\end{figure*}

In Figs. \ref{fig:high}, \ref{fig:zoom}, \ref{fig:mid} and \ref{fig:low} we show images of Nest200047 at all observed frequencies, including both the LOFAR frequencies (53 MHz and 144 MHz), already published in \cite{brienza2021}, and the new higher frequencies presented here for the first time (namely 400 MHz, 700 MHz, 815 MHz, 1280 MHz and 1518 MHz). For each band, we show three images (two at 53 MHz) at different angular resolutions to highlight the variety of structures on different scales (see Table \ref{tab:imageparam}). We note that all observations, except for those with uGMRT band-4 and VLA, are sensitive to the emission on scales of the entire source equal to $\sim$17 arcmin, which corresponds to $\leq$100$\lambda$. This implies that the decrease of large-scale, diffuse emission observed moving from low to high frequencies is mostly driven by the steep spectrum of the emission.

In the following text, we refer to these images as high-, mid- and low- resolution images. The main morphological features are labelled following \cite{brienza2021} and more labels are added for the secondary features as well here (see Fig. \ref{fig:presentation}). A zoom-in on the central region at the highest available resolution is also presented in Fig.~\ref{fig:zoom}.

The brightest structures in the entire source are the inner jets (marked as A), and the two pairs of inner bubbles (marked as B1/B2 and C1/C2, respectively), which are nicely detected at all frequencies. The mean value of surface brightness drops significantly when moving from region A to region C in all images. For example, in the high-resolution 144-MHz image we measure a drop of approximately a factor ten from A to C, with typical values of $\sim$15 mJy $\rm beam^{-1}$ in region A, $\sim$5 mJy $\rm beam^{-1}$ in region B and $\sim$1.5 mJy $\rm beam^{-1}$ in region C.

Interestingly, the higher resolution and sensitivity offered by uGMRT and VLA allow us now to appreciate much more clearly that even in this central region the morphology of the radio emission is very complex (see Fig. \ref{fig:zoom}). 
The inner jets (region A) extend almost symmetrically for about 22 kpc and then show some bending towards East, probably along the line of sight, suggesting the plasma gets detached from the jet flow. Moreover, as already noted previously, in the innermost regions of the jets we detect two symmetric very compact components, which could either be interpreted as signatures of new outbursts or recollimation shocks within the the jet flow.

Moving outwards, we can see that while the northern bubble B2 appears to have quite a regular shape, the bubble B1 in the south turns out to be very distorted: it has a clear edge on the East, and it morphs out into a double strand of emission towards north-west (marked as F5 and F6), which were poorly detected in the previous LOFAR images, probably due to a combination of resolution and sensitivity. 

Beyond this, we can see that the whole inner jets/lobes (A+B) are overall surrounded by a web of very low surface brightness emission, rich in thin and bent filamentary structures, oriented in all directions and with sizes in the range $\sim$1.5-3 kpc ($\sim$4-8 arcsec), as highlighted by the dashed lines in the top-right panel of Fig. \ref{fig:zoom}.
We also note that the bubbles B1 and B2 and, even more, the bubbles C1 and C2 are misaligned with respect to the direction of the inner jets, most likely suggesting a drifting during their buoyant rise in the group atmosphere, possibly caused by the system's weather.

Moving even more to larger scales, we can see that at all the new frequencies we recover, at least partially, the complex array of filaments and diffuse emission previously observed with LOFAR at 53 MHz and 144 MHz, especially at low resolution. In particular, the main distinctive feature of Nest200047, i.e. the bright filament (D2) perpendicular to the central jets and bubbles, is detected up to 1518 MHz, with a physical extension varying from 350 kpc to 290 kpc when moving from low to high frequency. 

The other fainter features, including the upper region of the outer bubble (D3), the southern outer bubble (D1), the boxy-shape filament, and multiple minor filaments, are instead only detected at some frequencies and at some specific resolutions. In none of the new images, instead, we detect the large-scale diffuse emission on the West likely due to a combination of low surface brightness and a very steep spectrum. 

Overall, the most complete detection of the previously unknown features in the new images is obtained in the uGMRT image at 400 MHz and in the MeerKAT image at 815 MHz, especially at lower resolution. Here we can clearly observe the outermost bubbles D3 and D1 and most of the filaments. Moreover, both images confirm the presence of a bridge of emission connecting (at least in projection) the bubble C2 and the filament D2 in the north, on the one hand, and the bubble C1 and the emission D1 in the south, on the other hand.

Thanks to the high resolution of the new images ($\lesssim$5 arcsec) we can further constrain the width of the large-scale filaments already detected in \cite{brienza2021}. Filaments F1, F2, F3, F4 and the boxy filament have median sizes of 4-5 kpc and down to 2-3 kpc. If considered as a single structure, the brightest filament in the north D2 has a varying width across its length that varies in the range 4-15 kpc ($\sim$12-45 arcsec). However, interestingly, at the new available resolution the eastern side of D2 appears to be actually composed of two almost parallel filaments, the brightest one being to the top (in projection) and each having a width of $\sim$5 kpc.

\subsection{Spectral index}
\label{sec:spix}

In \cite{brienza2021} we presented a spectral index map of the source in the frequency range 53-144 MHz with a resolution of 25 arcsec (9 kpc), aiming at probing the entire visible emission and already showing the overall steep-spectrum nature of the observed emission. Here we present new spectral index maps, covering a much larger frequency range and at different resolutions, to investigate in as much detail as possible the properties of all the different features. 

\begin{figure*}[!htp]
\centering
\includegraphics[width=0.8\textwidth]{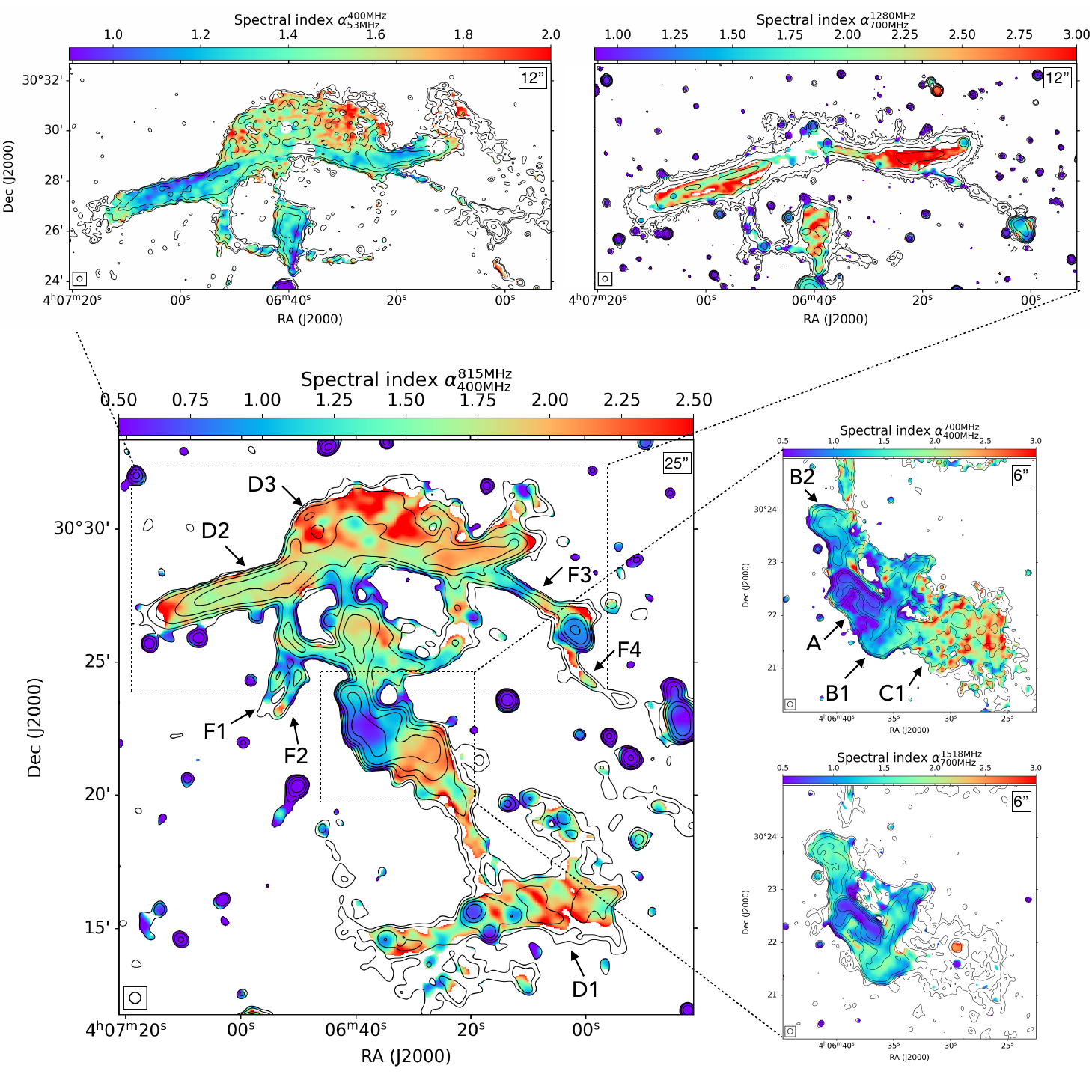}
\caption{Spectral index maps of Nest200047 as described in Sect. \ref{sec:spix}. \textit{Central panel}: spectral index map of the entire source between 400-815 MHz with a resolution of 25 arcsec. \textit{Top panels}: spectral index maps of the northern bubbles between frequencies 53-400 MHz (left), and 700-1280 MHz (right), with a resolution of 12 arcsec. \textit{Right panels}: spectral index maps of the central jets/bubbles between 400-700 MHz (top) and 700-1518 MHz (bottom) with a resolution of 6 arcsec. Black contours always trace the emission at the lowest frequency included in the map starting from $\rm 3\sigma$. }
\label{fig:spix}
 \end{figure*}
 
In particular, in Fig. \ref{fig:spix}, we show: 1) a spectral index map of the entire source between 400-815 MHz with a resolution of 25 arcsec (central panel); 2) two spectral index maps of the northern bubbles between 53-400 MHz and 700-1280 MHz with a resolution of 12 arcsec (top panels); 3) two spectral index maps of the central jets/bubbles between 400-700 MHz and 700-1518 MHz with a resolution of 6 arcsec. 

We note that, by design, the used telescopes have different UV coverage, making it impossible to achieve identical sensitivity across all relevant scales. To minimize these sensitivity differences and ensure consistent recovery of the radio emission from the same spatial scales while maintaining image good sensitivity, for our spectral analysis we created all images using a common Briggs weighting scheme with a Robust parameter of -0.5.  In this way, the detectability of any emission mainly depends on the thermal noise of the observation and the emission spectral shape. The images were smoothed to the same resolution, and identical UV cuts were applied at the shortest baselines for each pair/triplet of datasets (100$\lambda$ for the 25 arcsec map and 200$\lambda$ for the other two maps, driven by the uv-min of the uGMRT band-3 and band-4 datasets, respectively). Only pixels above 3$\sigma$ in all images involved were considered.

The spectral index maps between two frequencies were computed using the standard expression:

\begin{equation}
\rm \alpha = log_{10}(S_{\nu_1}/S_{\nu_2})/log_{10}(\nu_1/\nu_2),
\label{eq:spixerr}
\end{equation}

where $\rm S_{\nu_1}$ and $\rm S_{\nu_2}$ are the surface brightness values at frequencies $\rm \nu_1$ and $\rm \nu_2$, respectively.

Uncertainties on the spectral index were computed with the standard formula:

\begin{equation}
\rm \Delta\alpha = \frac{1}{ln\frac{\nu_1}{\nu_2}}\sqrt{\left(\frac{\Delta S_{\nu_1}}{S_{\nu_1}}\right)^2+\left(\frac{\Delta S_{\nu_2}}{S_{\nu_2}}\right)^2} ,
\label{eq:spixerr}
\end{equation}

\begin{figure*}[!htp]
\centering
\includegraphics[width=0.7\textwidth]{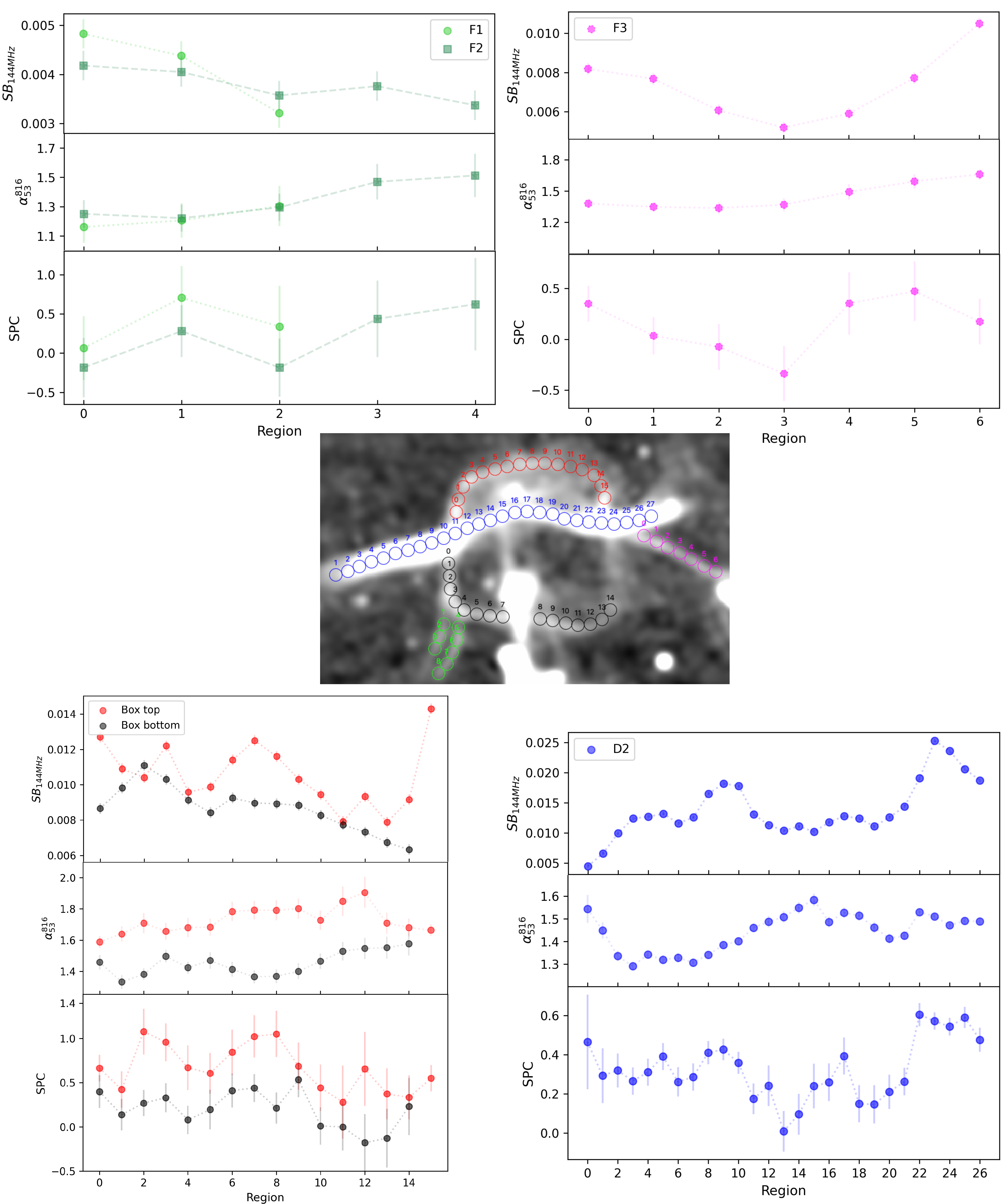}
\caption{Profiles of surface brightness (in mJy/beam), spectral index and spectral curvature across the main filamentary features observed in Nest200047. Images at 53, 144, 400 and 816 MHz with 25 arcsec resolution have been used for the measurements. The regions used are shown in the central panel and have a size equal to the beam size. Systematic uncertainties in the flux-scale were not included in the error bars as they do not affect the measured trend.
The surface brightness is measured at 144 MHz, the spectral index is computed in the between 53 MHz and 816 MHz and the spectral curvature is defined as SPC=$\rm\alpha_{400}^{816}$-$\rm\alpha_{53}^{144}$.}
\label{fig:profiles}
 \end{figure*}

\begin{figure*}[!htp]
\centering
\sidecaption
\includegraphics[width=0.6\textwidth]{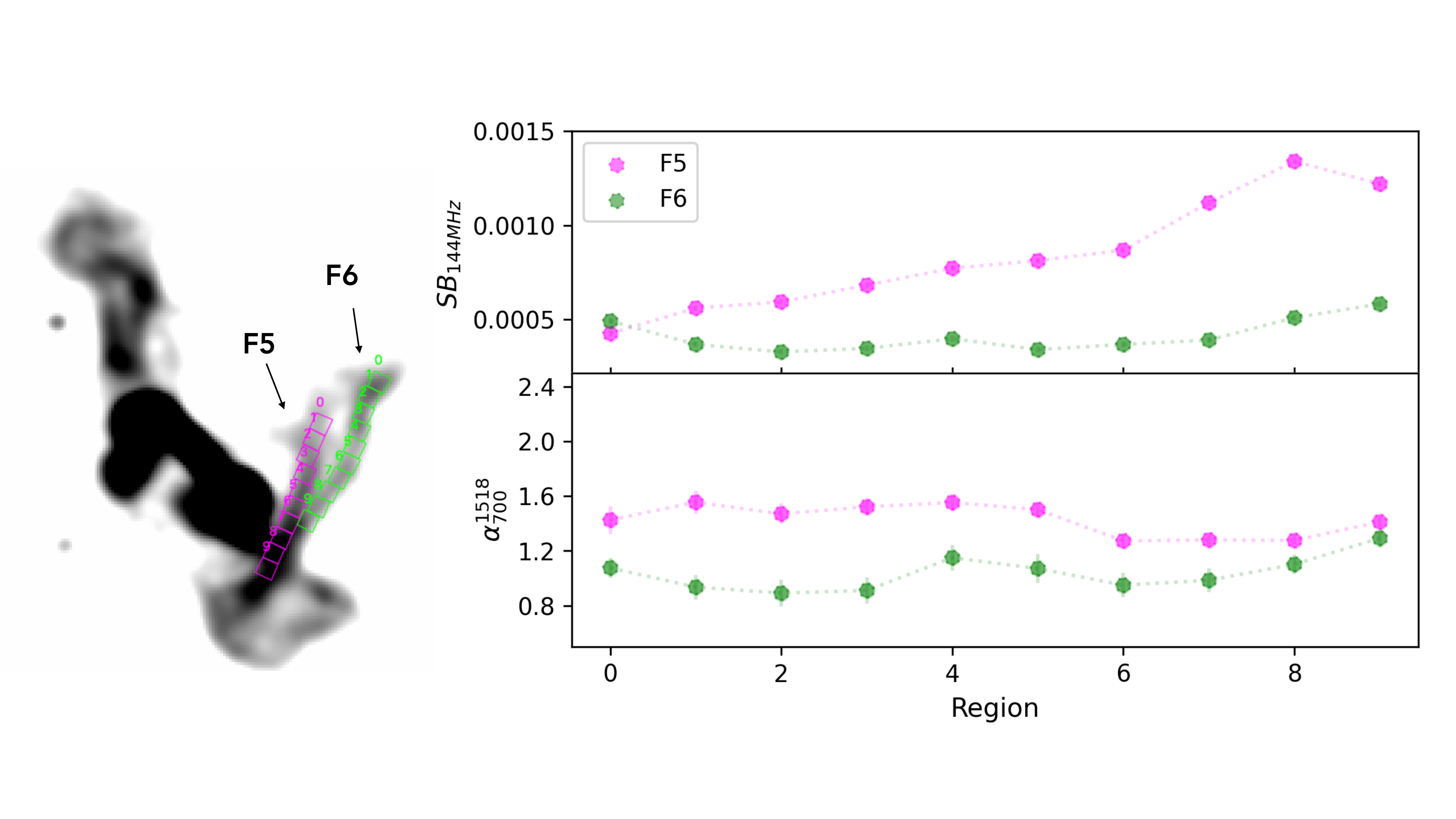}
\caption{Profiles of surface brightness (in mJy/beam) and spectral index across the two filamentary features, F5 and F6, observed in the inner region of Nest200047. Images at 700 MHz and 1518 MHz with 6 arcsec resolution have been used for the measurements. The regions used are shown in the left panel and have a size equal to the beam size. Systematic uncertainties in the flux-scale were not included in the error bars as they do not affect the measured trend. }
\label{fig:profiles-inner}
 \end{figure*}

 \begin{figure*}[!htp]
\centering
\includegraphics[width=0.7\textwidth]{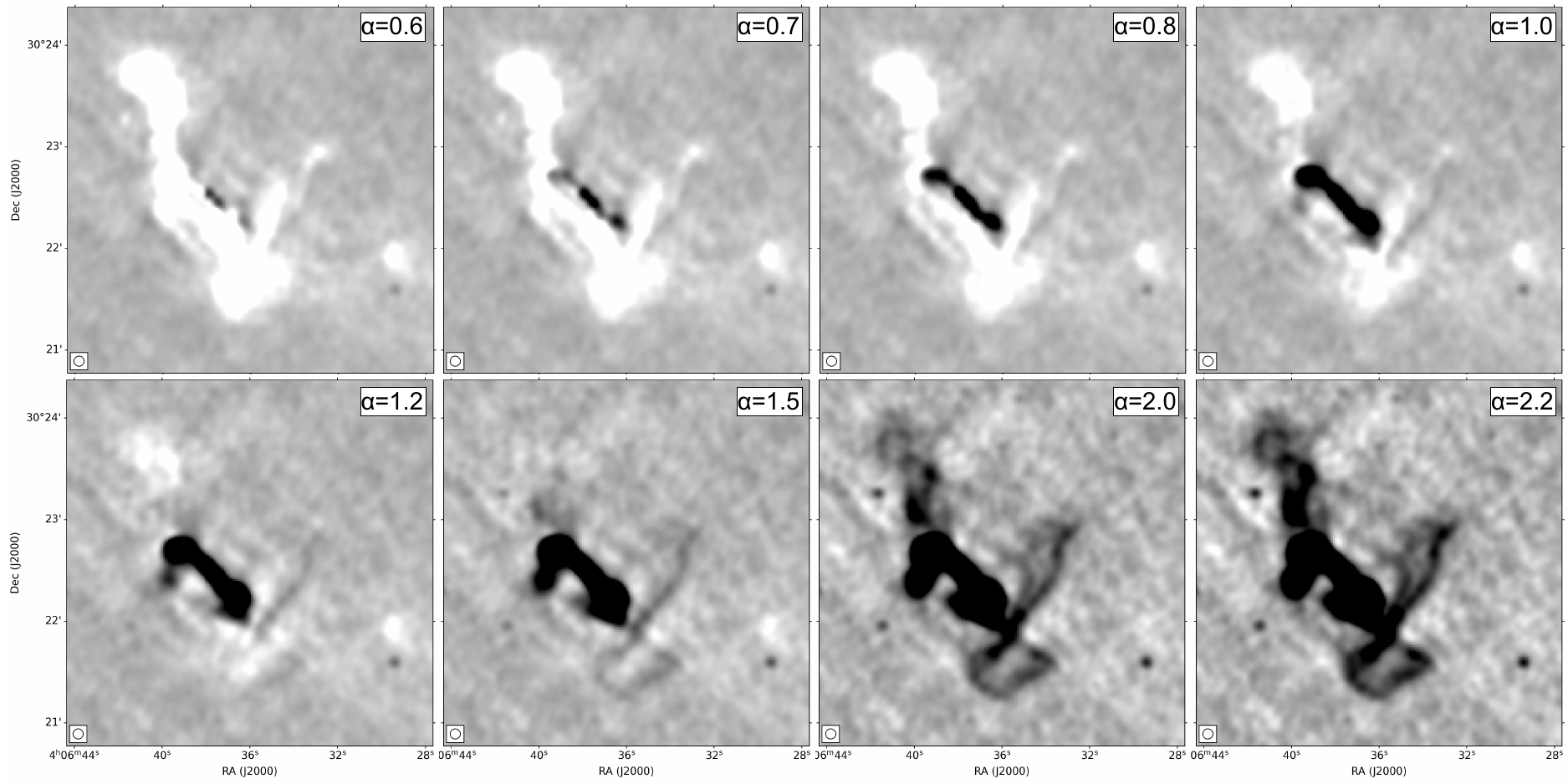}
\caption{Spectral tomography of the inner lobes in the frequency range 400-1518 MHz at 6 arcsec resolution. White emission is steeper than the reference $\alpha$ value reported in the top-right corner of each panel, while black emission is flatter.}
\label{fig:tomo1}
 \end{figure*}

 \begin{figure*}[!htp]
\centering
\includegraphics[width=0.7\textwidth]{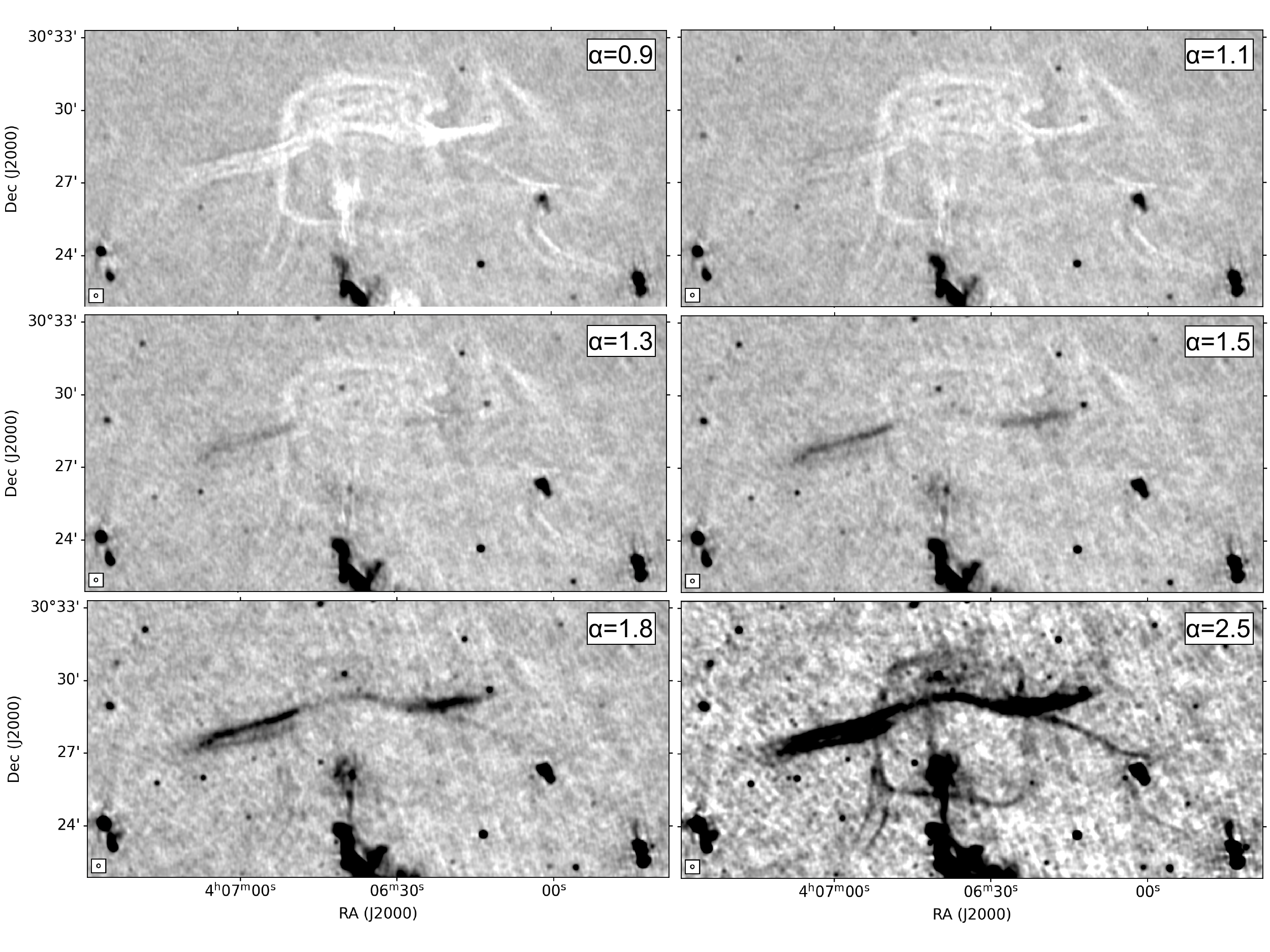}
\caption{Spectral tomography of the northern large-scale emission in the frequency range 144-400 MHz at 9 arcsec resolution. White emission is steeper than the reference $\alpha$ value reported in the top-right corner of each panel, while black emission is flatter.}
\label{fig:tomo2}
 \end{figure*}

where $\rm\Delta S_{\nu_1}$ and $\rm\Delta S_{\nu_1}$ are the surface brightness uncertainties computed by combining in quadrature the flux density scale calibration error ($\Delta S_c$) and the image rms noise ($\sigma$). The flux density scale calibration error is assumed to be 10\% for all images, except for the VLA image where it is set to 5\%, following the literature. Spectral index uncertainties maps are shown in Appendix \ref{appendix1}.

As we can appreciate from Fig. \ref{fig:spix}, in the map at 25 arcsec between 400 and 815 MHz we can see a large portion of the extended emission, including the filaments F1, F2, F3, F4, the boxy filament, the bubble D1 and the bridge of emission connecting D1 with C1. Overall, we can see that in this frequency range the spectral index of the various different components is systematically steeper than what observed at LOFAR frequencies \citep{brienza2021}, consistent with a curved spectrum. The steepest values (up to $\rm \alpha_{400}^{815}$=2.5) are measured in the upper region of the bubble D3 and in D1, while the flattest values ($\rm\alpha_{400}^{815}=0.6\sim1$) in correspondence of the inner jets and bubbles (A and B). The main filament D2 shows values in the range $\rm\alpha_{400}^{815}=1.6\sim2$, with the Western region being steeper than the Eastern one. The bubble C1 shows an average value of $\rm\alpha_{400}^{815}\sim2$, which is steeper than what observed in the specular bubble C2 having $\rm\alpha_{400}^{815}\sim1.5$. The boxy region and the smaller filaments F1, F2 and F3 have the overall flattest spectral index within the large-scale emission with values in the range $\rm\alpha_{400}^{815}=1\sim1.5$.

It is interesting to notice that at higher resolution (see upper panels of Fig. \ref{fig:spix} at 12 arcsec) a clear spectral gradient across the width of the main filament D2 is present. In particular, the northern regions (especially to the west), show a flatter spectral index (with $\rm\alpha_{53}^{400}\sim1$ and $\rm\alpha_{700}^{1280}\sim2$) than the southern regions (with $\rm\alpha_{53}^{400}\sim2$ and $\rm\alpha_{700}^{1280}\sim3$). 

A hint of this trend was already found in the low-resolution (25 arcsec) spectral index map between 53-144 MHz published in \cite{brienza2021}, and was interpreted as a possible signature of plasma compression produced by a shock moving from south to north, consistent with being initiated by the recurrent AGN outbursts of the central elliptical galaxy MCG+05-10-007.

However, as already mentioned in Sect. \ref{sec:morpho}, when observed at high resolution (see Fig. \ref{fig:high}), the Eastern side of the filament D2 appears to be actually split into two almost-parallel substructures. This may suggest that the observed spectral index trend is likely only produced by the proximity of these two features with different spectral properties.

Finally, in the spectral index maps at 6-arcsec resolution (Fig. \ref{fig:spix}, right panel) we can better appreciate the curvature increase when moving from region A to regions B and C. In particular, region A is the only one where the spectral index remains flatter ($\leq$1) up to 1518 MHz, consistent with ongoing particle acceleration. The surrounding bubbles B and the close-by filamentary emission instead steepen on average from a spectral index of $\rm\alpha_{400}^{700}\sim1-1.3$ to a spectral index of $\rm\alpha_{700}^{1518}\sim1.3-1.6$.

To further investigate the spectral distribution along the large-scale filamentary structures we produced spectral profiles as shown in Fig. \ref{fig:profiles}. For all filaments we computed the spectral index over the full available frequency range $\rm\alpha_{53}^{816}$ and the spectral curvature defined as SPC=$\rm\alpha_{400}^{816}$-$\rm\alpha_{53}^{144}$. The surface brightness at 144 MHz is also shown as a reference in the top panel of each quadrant.

For these plots we used images at a common resolution of 25 arcsec to best recover these low-surface brightness features. Only images at frequencies equal to 53 MHz, 144 MHz, 400 MHz and 816 MHz were included, where the filaments are detected at a S/N>5. The boxes used are shown in the central panel of the same figure and have a size equal to the beam size.

Overall, all filaments show a steep ($\rm\alpha_{53}^{816}$ in the range 1$\sim2$) and curved (SPC up to 1) spectrum over the available frequency range, without any strong and systematic spectral index gradient, as expected in the case of particle cooling along them as for example observed along the tails of bent radio galaxies (e.g. \citealp{sebastian2017}). 
Particularly impressive is the spectral index along the main filament D2 (blue boxes), whose values stay in the range $\rm\alpha_{53}^{816}$=1.2-1.6 for over 350 kpc. As already observed in the spectral index maps a moderate flattening moving from East to West is present. 

Similarly to the outer filaments, we produced spectral profiles for the two main filaments, F5 and F6, located in the inner region of the source. Images at 700 MHz and 1518 MHz with 6 arcsec resolution have been used for the measurements. As it is shown in Fig. \ref{fig:profiles-inner} and in agreement with the spectral index map (see Fig. \ref{fig:spix}), both filaments show a steep spectrum, with average values $\rm\alpha_{700}^{1518}(F5)$=1.5 and $\rm\alpha_{700}^{1518}(F6)$=1, respectively. F5 is systematically steeper than F6 and, also in this case, both filaments show a quite constant spectral trend throughout their length.

\label{sec:radio}

\subsection{Spectral tomography}
\label{sec:tomo}

In such a complex source, plenty of thin, possibly overlapping features, spectral index maps may not be sufficient to fully appreciate all spectral shades across the different structures. To overcome this limit, we used the `spectral tomography' technique first introduced by \cite{katzstone1997} and later used in a number of works on different astrophysical sources (e.g. \citealp{katzstone1997b, crawford2001, delaney2002, gizani2003, mckinley2013, rajpurohit2021}). This is useful for visualising features with different spectral indices and superimposed structures. The idea behind this technique is to subtract out of an image all features with a particular spectral index value so that only features with a flatter or steeper spectral index remain visible. To do so, we subtracted from an image at a frequency $\nu_1$ a second image at frequency $\nu_2$ scaled by a specific spectral index value $\alpha_i$, in the following way:

\begin{equation}
    I(\alpha_i)=I_{\nu_1}-\left(\frac{\nu_1}{\nu_2}\right)^{-\alpha_{i}}\cdot I_{\nu_2} .
\end{equation}

In Fig. \ref{fig:tomo1} and \ref{fig:tomo2} we present a gallery of the most representative tomography maps for the central and northern, large-scale region of Nest200047, respectively. For the central region we used images at $\nu_1$ = 144 MHz and $\nu_2$ = 400 MHz with 6-arcsec resolution, while for the northern region we used images at $\nu_1$ = 400 MHz and $\nu_2$ = 1518 MHz with 9-arcsec resolution. The spectral index $\alpha_i$ used in each map is shown in the top right corner of each panel.
In these images all structures having $\alpha_i$ are fully subtracted and thus are invisible. Instead, all structures that are steeper or flatter than $\alpha_i$ are under- or over-subtracted and thus appear lighter or darker than the background, respectively.

Fig. \ref{fig:tomo1} shows a zoom-in on the central region of Nest200047, where the inner jets and surrounding bubbles/filaments are visible (components A, B1 and B2) and their different spectral indices can be appreciated. Overall, as expected and consistently with the spectral index map presented in Sect. \ref{sec:spix}, the spectral index steepens moving from the inner jet toward the external regions. However, many more details can be observed here. 

In the top-left panel, we can see that most of the emission is white, implying it has a spectral index higher than $\alpha$=0.6. The only exception to this are the two very compact components in the innermost region of the active jet, which likely represent regions where particles are being accelerated. In the other panels on the top row of Fig. \ref{fig:tomo1} we can then see how the plasma gets steeper and steeper as it expands both along and across the jets (component A). This trend is consistent with observations of Fanaroff-Riley class I \citep{fanaroff1974} radio galaxies \cite{heesen2018}. From $\alpha$=1 we can start seeing that after having reached the southern and northern tip of component A, the jet flow bends towards East (in projection).

The tomography clearly confirms that the two main filaments F5 and F6 have different spectral indices, with F5 being steeper than F6. 
Interestingly, F6 seems to show a spectral index gradient across its width. It can be noted that the Western side of F6 becomes darker first that the Eastern side. To the south, the filament F6 seems to bend and continue tracing the full edge of the bubble B1. 

Overall, the bubbles B1 and B2 get completely blended with the background at $\alpha\sim$1.5 suggesting this is the average spectral index of their spectrum, consistent with the spectral index maps. However, it is interesting to note that the edge of bubble B1 is flatter.

In Fig. \ref{fig:tomo2} we show instead a zoom-in on the northern large-scale emission of Nest200047 (regions C2, D2 and D3 and connected filaments). The emission is entirely white in the top-left panel, implying it has a spectrum steeper than $\rm \alpha_i$=0.9. Starting from $\rm \alpha_i\sim$1.1 some of the emission starts to disappear, especially the Eastern region of the main filament D2. 

The tomography maps show more clearly that the eastern side of D2 has a flatter spectral index than the western side and that the eastern side is actually composed of two almost parallel substructures with different spectral properties, the upper one being flatter than the bottom one.

\subsection{Surface brightness vs spectral properties}
\label{sec:bright}

\begin{figure*}[!htp]
\centering
\includegraphics[width=0.8\textwidth]{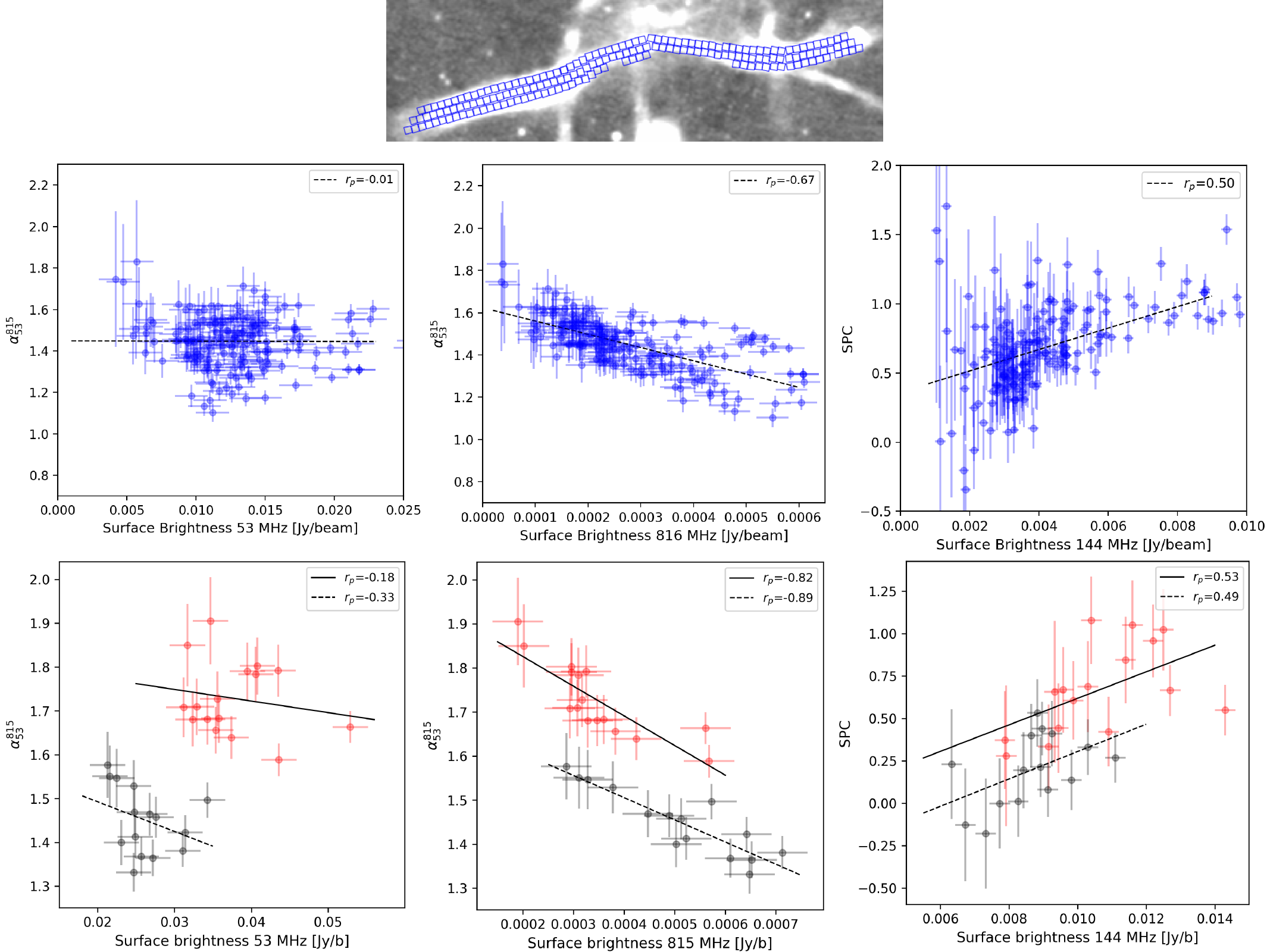}
\caption{Radio surface brightness vs spectral index (left and middle panels) and spectral curvature (right panel) for filaments D2 (blue) and box bottom (black) and top (red). The radio spectral curvature is computed as SPC=$\alpha_{400}^{816}-\alpha_{53}^{144}$. The regions used for D2 are shown in the top panel of the figure, while the regions used for the boxy filament are the same as in Fig. \ref{fig:profiles}. A power-law fit to the data is shown with a line and the respective Pearson coefficient $r_p$ is shown in the top-right corner of each panel.}
\label{fig:bright}
\end{figure*}

We investigated the presence of any correlation between radio surface brightness $S$ and spectral shape as shown in Fig.~\ref{fig:bright}. In particular, we show correlations with the spectral index $\alpha_{53}^{815}$ and spectral curvature SPC=$\alpha_{400}^{816}-\alpha_{53}^{144}$.
Measurements for D2 were performed on the 12-arcsec images using the blue boxes shown in the same figure, while measurements for the boxy region were performed from the 25-arcsec images using the boxes presented in Fig. \ref{fig:profiles}. For each plot a power law was fitted to the data and the goodness of the fit was evaluated with the Pearson coefficient $r_p$. 

The analysis in general shows that all three filaments have the same behaviour: a strong ($r_p$=0.67-0.89) negative correlation between $\alpha_{53}^{815}$ and surface brightness at 815 MHz, which becomes weaker and weaker as it moves to the lower frequency, and a positive correlation between SPC and surface brightness at all frequencies. In other words, the brightest regions at high frequency have more curved spectra but with more pronounced spectral curvature with respect to the fainter regions.

\subsection{Colour-colour plots}
\label{sec:color}

The spectral index and tomography maps presented in the previous sections clearly show the spectral index trends throughout the source. However, they are not ideal to visualize the curvature of the spectrum across the broad frequency range available. To do this, we employ colour-colour plots ($\alpha_{\rm  low}$ vs $\alpha_{\rm high}$), first introduced by \cite{katzstone1993, katzstone1997}.

In these plots, all points located along the bisector line with $\rm \alpha_{low} = \alpha_{high}$ represent regions of plasma with a power-law spectrum, while points located above or under the bisector line indicate a concave or convex spectra, respectively. In the common hypothesis that the spectrum at injection ($\rm t_0$) follows a power law and then steepens with time due to radiative cooling, it is expected that a single population of particles is located on the bisector at $\rm t_0$ (at the injection value $\rm \alpha_{inj}$) and then, with time, to move along a trajectory below the bisector line. 
The exact shape of the trajectory depends on the ageing model, the most common ones being the JP \cite{jaffe1973}, KP \cite{kardashev1962} and CIOFF \cite{komissarov1994} models. In particular, the JP and KP models describe the radiative losses of a single-injection particle population through synchrotron emission and inverse-Compton scattering with the CMB assuming rapid isotropisation or constancy of the particles' pitch angle, respectively (i.e. the angle between the velocity vector and the magnetic field). The CIOFF model instead considers an extended period of constant injection with given start and end times. All these models are derived for cooling in homogenous magnetic fields. We also note that the trajectories along the colour-colour plots are insensitive to the magnetic field strength and to the presence of adiabatic compression or expansion in the plasma. These factors can only cause a shift up or down along the ageing trajectory but do not cause a deviation from it. 

We note that for a given single-injection model the injection can in principle be inferred by extrapolating the trajectory in the colour-colour plot up to the bisector line. Finally, the area of the plot above the bisector line can be populated by regions where particle populations with different ages/slopes are mixed (or overlap in projection) or a fraction of particles has been re-accelerated, leading to the inverted spectral curvature.

\begin{figure*}[!htp]
\centering
\includegraphics[width=0.8\textwidth]{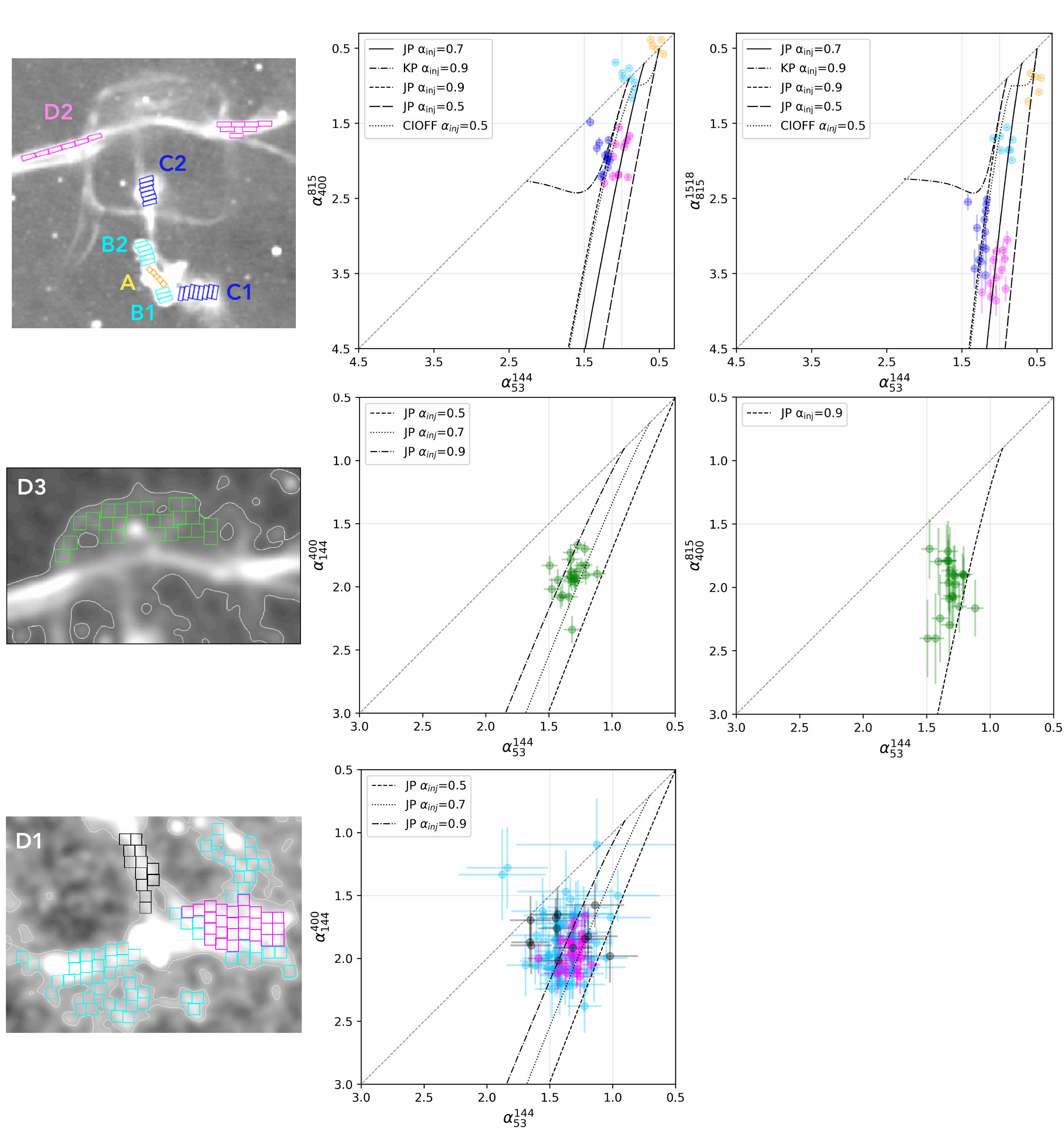}
\caption{Colour-colour plots for different regions across Nest200047 and over different frequencies. The top row shows $\rm \alpha_{53}^{144}$ vs $\rm \alpha_{400}^{815}$ and $\rm \alpha_{53}^{144}$ vs $\rm \alpha_{815}^{1518}$ for regions D2 (magenta points), C1/C2 (blue points) and B1/B2 (cyan points) and A (yellow points). Images at 12-arcsec resolution were used for the measurements. The boxes used (one-beam area for region A and three beam-area for the other regions), colour-coded according to the different regions, are shown in the top-left panel. 
The middle row shows $\rm \alpha_{53}^{144}$ vs $\rm \alpha_{144}^{400}$ (middle) and $\rm \alpha_{53}^{144}$ vs $\rm \alpha_{400}^{815}$ (left) for D3 (green points). The bottom row shows $\rm \alpha_{53}^{144}$ vs $\rm \alpha_{144}^{400}$ for the southern large-scale region, including D1, the bridge and the diffuse emission. For the plots shown in the middle and bottom row, images at 25-arcsec were used for the measurements. The boxes used (one-beam area), colour-coded according to the different regions, are shown in the left panels of each row, respectively. In all plots the dashed line represents the bisector line (where $\rm \alpha_{high}=\alpha_{low}$), the other lines represent various ageing models as a reference.}
\label{fig:cc-all}
 \end{figure*}

In Fig. \ref{fig:cc-all} we show colour-colour plots for different regions of Nest200047. In particular in the top panel we analysed region A, B1/B2, C1/C2 and D2, in the middle panel region D3 and in the bottom panel region D1, the bridge connecting D1 and C1, and some diffuse emission on the SE.

In the top panel, we show two colour-colour plots probing the low and the high frequency part of the spectrum, respectively (53-144 vs 400-815 MHz on the left and 53-144 vs 815-1518 MHz on the right). For these we used the intermediate resolution maps presented in Sect. \ref{sec:spix} with a beam of 12 arcsec. The boxes used to produce the plot (see left panel) have a size equal to 12 arcsec $\times$ 12 arcsec for region A and 12 arcsec $\times$ 36 arcsec for the other regions, and were all drawn only in areas with S/N>3 in all maps (with the main limitation coming from the VLA map at 1518 MHz) and are coloured based on the region probed (magenta for the filament D2, blue for the bubbles C1/C2 and cyan for the bubbles B1/B2). Various curves describing ageing models with different $\rm \alpha_{inj}$, are also overlaid in the plot as a reference.

From the figure we can see that region A is consistent in both plots with a JP model with $\rm \alpha_{inj}$=0.5. The trajectory followed by the cyan points (lobes B1/B2), instead, is more complex. At low frequency, they touch the bisector line at $\rm \sim\alpha$=0.9, but then they significantly deviate from the JP line with $\rm \alpha_{inj}$=0.9, showing less curvature at low frequency than what expected by the model. On average, they show a spectral curvature of about SPC=$\rm \alpha_{high}-\alpha_{low}\sim$0.7.

The plasma in the regions C1/C2 (blue points) and D2 (magenta points) is clearly much more curved than in B1/B2 and A at both low and high frequencies, with spectral curvature values up to SPC$\sim$0.8 at low frequencies (middle panel), and up to SPC$\sim$2.5 at high frequencies (right panel). This extreme spectral curvature at high frequency is clearly inconsistent with a KP model, as shown in Fig. \ref{fig:cc-all}, right panel. The trajectory of the blue regions appears to be nicely consistent with a JP model with a quite steep injection index $\rm \alpha_{inj}\sim$0.9. The magenta regions (D2), instead, show a larger scatter and, on average, a better fit with a JP curve with a lower injection index $\rm \alpha_{inj}\sim$0.7.

\begin{figure*}[!htp]
\centering
\includegraphics[width=0.85\textwidth]{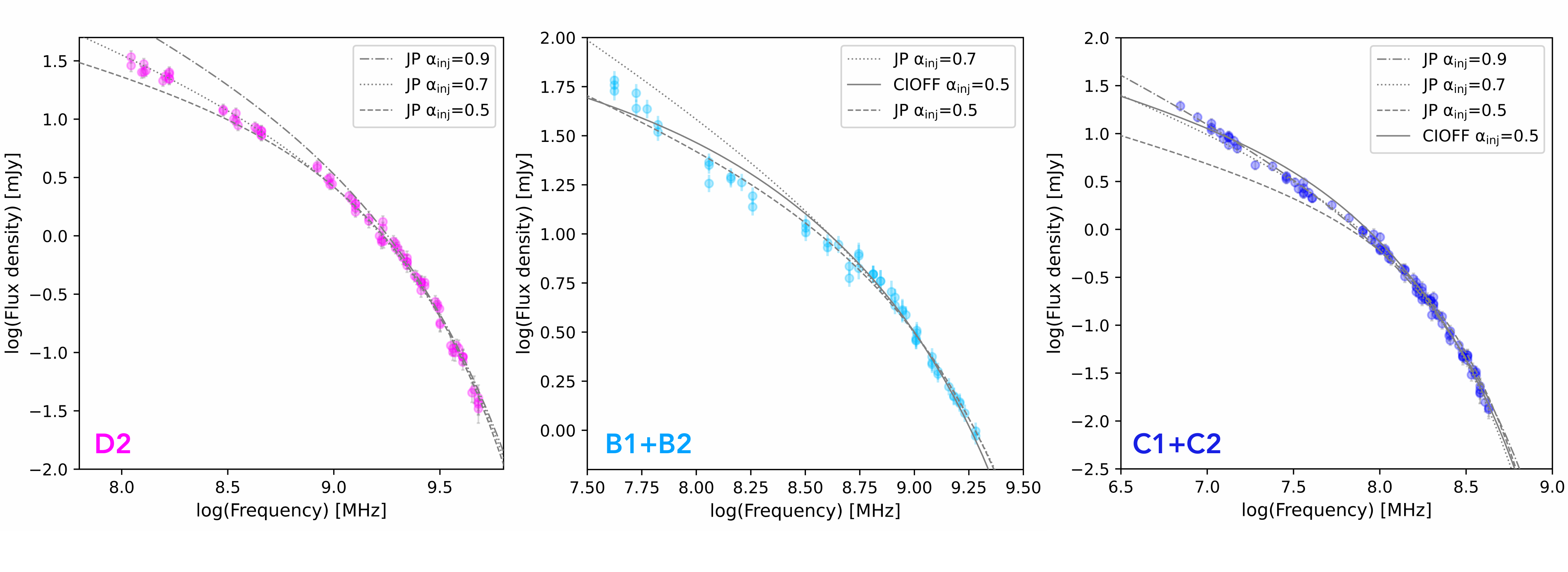}
\caption{Global spectra in the frequency range 54-1518 MHz for three different regions of the source Nest200047 (D2, C1/C2 and B1/B2) 
including flux densities at 53 MHz, 144 MHz, 400 MHz, 700 MHz, 815 MHz, 1280 MHz, and 1518 MHz. Blue points represent boxes drawn in region C1/C2, magenta points in region D2 and cyan points in B1/B2, as shown in Fig. \ref{fig:cc-all}. Lines represent various spectral models as a reference.}
\label{fig:global}
\end{figure*}

We note that, since the points of regions C1/C2 and D2 (blue and magenta, respectively) never touch the bisector line the injection is not constrained. In this case a degeneracy is present between the JP models with higher injection index values and the CIOFF model with lower injection index values. In Fig. \ref{fig:cc-all}, we show the curve of a CIOFF model with $\rm \alpha_{inj}\sim$0.5 and $\rm t_{on}$=60 Myr (dotted line), which may represent a plausible duration of active phases of an AGN. However, other combinations of higher $\rm \alpha_{inj}$ and lower $\rm t_{on}$ could also reproduce the data, also keeping in mind that bubbles A, B, C, D possibly originate from different activity phases with different on-times.

In the middle row of Fig. \ref{fig:cc-all} we show two colour-colour plots for region D3: $\rm \alpha_{53}^{144}$ vs $\rm \alpha_{144}^{400}$ (middle) and $\rm \alpha_{53}^{144}$ vs $\rm \alpha_{400}^{815}$ (right). For these we used images at 25 arcsec resolution to best recover the diffuse emission and we restrict the analysis to the frequencies where the emission is detected with S/N>3. The boxes used for the measurements (with 25-arcsec side) are shown in the left panel. As the plots show, at low-frequency (middle panel) the points are consistent with a JP model with $\rm \alpha_{inj}\sim$0.7-0.9, while at higher frequency (right panel) the points deviate to the left and are only marginally consistent with a JP model with $\rm \alpha_{inj}\sim$0.9.

Finally, in the bottom row we show one colour-colour plot of the southern large-scale structures. Also for this we use images at 25 arcsec resolution and boxes with 25 side as shown in the left panel. In general, most points lie below the bisector line but show large scatter and large error bars, making it difficult to find a match with a single radiative model. On average, the magenta points (which correspond to the brightest region of the D1 structure) seem to be consistent with a JP with $\rm \alpha_{inj}\gtrsim$ 0.7, similar to what found for D2/D3 in the north. 

In conclusion, this analysis shows that it is hard to reconcile all regions to the same spectral shape, as it is found for standard radio galaxies (e.g. \citealp{shulevski2017}.
While in principle, there is no reason to believe that different outbursts must accelerate populations of particles with the same injection index and that the injection phase has always the same duration, in such a complex system a different spectral shape may also be driven by different fluctuations of magnetic field or particle reprocessing.

\subsection{Global spectrum}
\label{sec:global}

To further investigate the spectral shape of the emission over the full available bandwidth we applied the `shift-technique" proposed by \cite{katzstone1993, katzstone1997}. The idea behind this is that, if the spectral shape difference between two particle populations is only due to different energy losses and magnetic fields, one should be able to match the two spectra to a common `global spectrum' by shifting one spectrum with respect to the other in both frequency and intensity (i.e. in the log~(I)-log~($\rm \nu$) space). In particular, shifts along the frequency axis are related to $\gamma^2B$ and along the intensity axis are related to $N_TB$, where $N_T$ is the total number of relativistic electrons.

To do this we used images at 12 arcsec resolution and the same boxes created for the colour-colour plot in Sect. \ref{sec:color}. In particular, for each region (D2, C1/C2 and B1/B2) we tried to align the spectrum of each box to create a global spectrum using as a reference the best models identified in the colour-colour plots. 

In Fig. \ref{fig:global} we show the result of this alignment procedure. From this we can clearly see that the spectra from all magenta boxes (D2) can be very well aligned to a JP model with $\rm \alpha_{inj}=$0.7, while the spectra from all blue boxes (C1/C2) can be very well aligned to a JP model with $\rm \alpha_{inj}=$0.9. For regions C1/C2 (left panel) we also overlay the CIOFF model with $\rm \alpha_{inj}=$0.5, which based on the colour-colour spectrum seemed to reproduce the curve as well as the JP model with $\rm \alpha_{inj}=$0.9. From this we can appreciate that there is a marginal difference between the two models, with possibly a slight preference for the JP model. If true, this would suggest that regions C1/C2 do really have a steeper injection index than regions B1/B2 rather than being an effect of superposition of different particle populations described by a CIOFF model.

For the cyan boxes (B1/B2), instead, the result of the alignment is much poorer for all the models considered (including both JP and CIOFF), especially at the lowest frequencies. This may hint at some more complex physical condition of the plasma, which may not be surprising considering that both lobes show at high resolution numerous filamentary substructures (see e.g. Fig. \ref{fig:tomo1}).

We note that we tested whether the results are sensitive to the size of the boxes used, but within the uncertainties, we always found the same trends. In conclusion, aligning the spectra of all regions to a single global spectrum is not possible, as already suggested by the colour-colour analysis presented in Sect. \ref{sec:color}. This again suggests that the properties of the different pairs of bubbles and filament D2 and/or the particle acceleration efficiencies over different scales, and so over different outbursts, are not constant.

\subsection{Spectral ages}
\label{sec:age}

The broad frequency coverage and large extension of the source allow us as well to perform a pixel-by-pixel spectral ageing analysis do derive the plasma age, as already done for other radio galaxies in the literature (e.g. \citealp{hardcastle2013, brienza2021, biava2021}). 

For this, we employed the software {\tt BRATS} (\citealp{harwood2013, harwood2015}), which derives modelled spectra by numerically integrating the equations that describe the radiative losses of the plasma. In particular, based on the results from the colour-colour plot and global spectrum analysis (see Sect. \ref{sec:color} and \ref{sec:global}) we used the JP model, which best reproduces the spectral curvature. 

We processed regions D3, C1/C2 and B1/B2 separately, to account for differences in injection index and magnetic field values. 
Regions A and D1 were not included in the analysis because of the small extent and small spectral coverage, respectively. For regions C1/C2 and B1/B2 we used the full set of images at 12 arcsec resolution, which gives a good compromise between having good spatial resolution and recovering the large scale emission, while for D3 we used images at

\begin{figure*}[!htp]
\centering
\sidecaption
\includegraphics[width=0.55\textwidth]{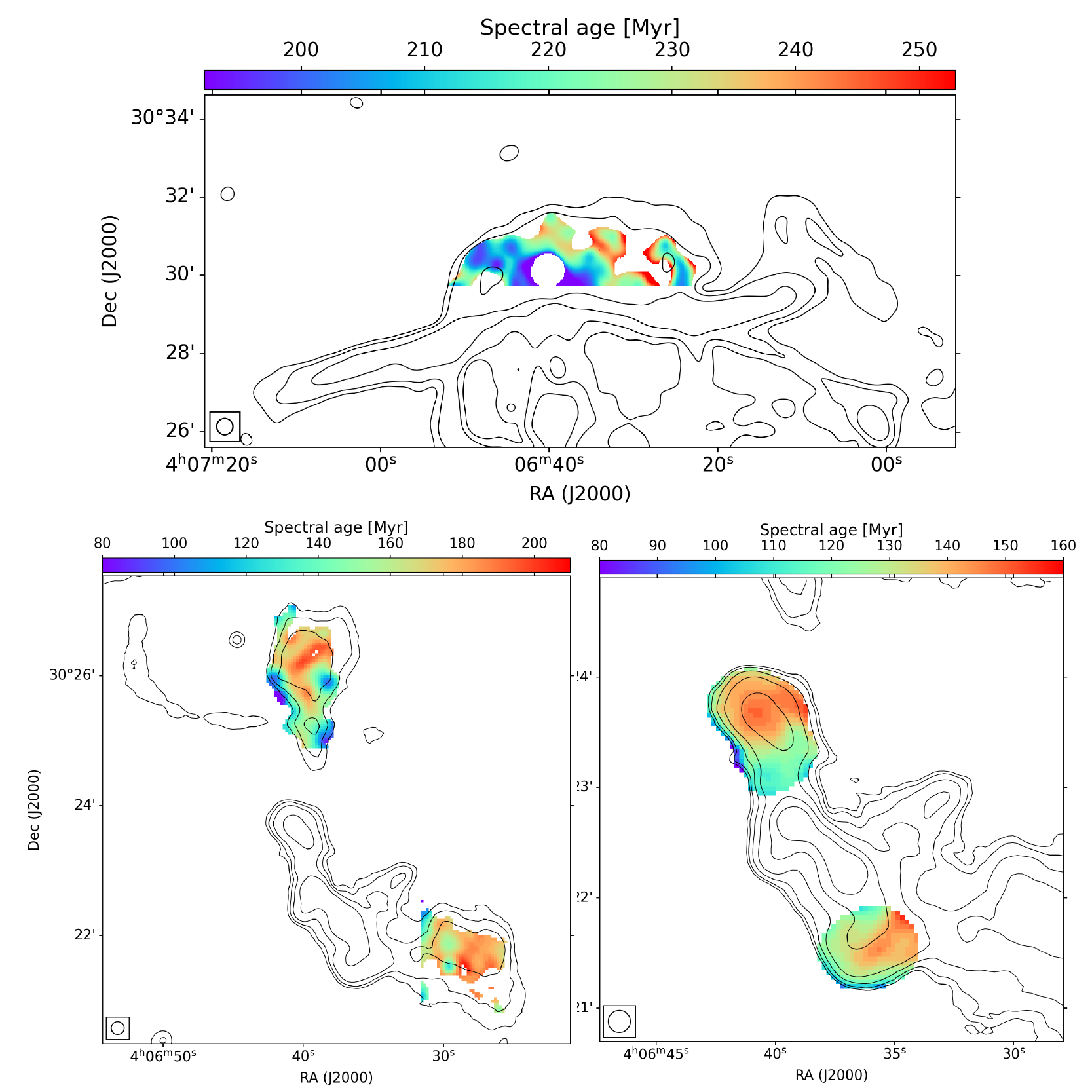}
\caption{Spectral age maps of different regions of Nest200047: D3 (top panel) at 25-arcsec resolution, and C1/C2 (bottom-left panel) and B1/B2 (bottom-right panel) at 12-arcsec resolution. Overlaid are contours of the LOFAR image at 144 MHz, at 25 and 12 arcsec, respectively. The full procedure followed to obtain the maps is described in Sect. \ref{sec:age}.}
\label{fig:age}
\end{figure*}

\noindent 25 arcsec in the frequency range 53-815 MHz, where the structure is recovered. In all cases only pixels with S/N>3 at all the frequencies were included in the analysis.

To establish the injection index we used the {\tt findinject} task. This performs a series of spectral fits for each pixel, varying $\rm \alpha_{inj}$ in the range 0.5 and 1 with a step of 0.05, and defines the best value based on the $\chi^2$. The best injection index value found for the different regions is $\rm \alpha_{inj}^{D3}$=0.9, $\rm \alpha_{inj}^{C1/C2}$=0.85 and $\rm \alpha_{inj}^{B1/B2}$=0.5, in good agreement with the values derived from the colour-colour plot and global spectrum analysis.

As for the magnetic field $B$, we assumed as a first simplified assumption, the value that returns the minimum radiative losses allowed for a plasma at a given redshift, equal to $\rm B_{min} \sqrt{3}\sim B_{CMB}$=1.95 $\rm \mu$G. On top of this, we considered the values presented in \cite{brienza2021} for D3 and C1/C2 based on minimum energy conditions equal to $B_{eq}^{D3}=2.8 \ \mu G$ and $B_{eq}^{C1/C2}$=4.2 $\mu G$, respectively. With the same approach we computed a value of $B_{eq}^{B1/B2}$=3.5 $\mu G$ for B1/B2. 

The spectral age maps obtained using $\rm B_{min}$ are shown in Fig.~\ref{fig:age}, with overlaid LOFAR 144-MHz contours. Considering all pixels for each region we find that in all cases the model cannot be rejected at the 68 percent significance. The median $\chi^2_{red}$ is 0.7 [0.2-7], 1.2 [0.08-7] and 1.8 [0.16-16] for D3, C1/C2 and B1/B2, respectively. The median error on the fit $\rm \Delta$t is 15 Myr [9-30 Myr], 5 Myr [1-10 Myr], and 4 Myr [1-10 Myr]. Examples of good spectral fits for the different regions are shown in Appendix~\ref{appendix2} as a reference.

We note that regions B1/B2 show the highest $\chi^2_{red}$, suggesting that the plasma in that region is not following the standard spectral shapes similarly to what observed with the global spectrum analysis. This is maybe not surprising given that from the images at high resolution (see Fig. \ref{fig:zoom}), it is clear that lobe B1 is constituted by twisted filaments, which might be tracing regions of compression and/or magnetic field enhancements.

Excluding regions right at the edges of the maps, which might be dominated by residual alignment artefacts, we find the following median values for the different regions: 220 Myr for D3, 160/170 Myr for C1/C2 and 130 Myr for B1/B2. These correspond to break frequencies of 430 MHz, 730 MHz and 1200 MHz for D3, C1/C2 and B1/B2 regions, respectively. The ages for the different regions are summarised in Table \ref{tab:ages}, where we also report values obtained using the minimum energy magnetic field $B_{eq}$, which are systematically lower, as expected for a higher magnetic field value.

We note that these are the ages computed only in the regions of the source where full spectral coverage (53-1518 MHz) is available at 12 arcsec resolution. There are clearly regions not recovered at all frequencies that likely have larger age values.

\begin{table*}[!htp]
\centering
        \small
 \caption{Summary of the best fit spectral ages for different regions of the source obtained through spectral ageing modelling as described in Sect. \ref{sec:age}. }
                \begin{tabular}{c c c c c c}
                \hline
                \hline
                & D3 & C1 & C2  & B1 & B2  \\
                & $\rm\dot{t}$ [$\rm t_{min}-t_{max}$] Myr  &$\rm\dot{t}$ [$\rm t_{min}-t_{max}$] Myr & $\rm\dot{t}$ [$\rm t_{min}-t_{max}$] Myr& $\rm\dot{t}$ [$\rm t_{min}-t_{max}$] Myr& $\rm\dot{t}$ [$\rm t_{min}-t_{max}$] Myr\\
                \hline
                \hline
                \\
                $\rm B_{min}$ &220 [170-260]  & 160 [115-190]  & 170 [120-190] & 130 [100-155]  & 130 [100-150] \\
                \\
                $\rm B_{eq}$ & 200 [180-250]& 130 [115-160] & 130 [90-150] & 115 [90-130] & 115 [90-130]\\
                \hline
                \hline  
                \end{tabular}
                
\label{tab:ages}
\end{table*}

\section{Discussion}
\label{sec:discussion}

\subsection{Jet duty cycle}
\label{sec:time}

Nest200047 represents arguably the clearest example of multiple AGN radio bubbles observed in a galaxy group. Therefore, the source is a perfect testbed for evaluating the timescales of the jet duty cycle.

As expected in a scenario where the non-thermal plasma is generated by the black hole at the centre of the BGG MCG+05-10-007 and moves outward, and consistent with the observed spectral index trends (see Sect. \ref{sec:spix}), the median age values based on simple radiative age modelling (see Sect. \ref{sec:age}) increase moving from the inner regions of the source toward the outskirts (see Table \ref{tab:ages}). In particular, the median ages are very similar for the pair of bubbles C1 and C2 (the same is true for B1 and B2 pair), supporting the interpretation that they are associated with the same outburst as suggested by their morphology, similar distance from the host, and spectral index value (see Sect. \ref{sec:spix}). 

To derive a first-order estimate of the duration of the jet activity that inflated a given bubble, $\rm t_{on}$, we can calculate the difference between the maximum and minimum age values derived within each bubble, as done in other works (e.g. \citealp{shulevski2017}). This provides an estimate of the jet activity times for the three different bubbles equal to $\rm t_{on}^{D3}$>90 Myr, $\rm t_{on}^{C1/C2}$=60 Myr and $\rm t_{on}^{B1/B2}$=50 Myr, all values consistent with other radio galaxies (e.g. \citealp{saikia2009, konar2012, konar2013, shulevski2017, brienza2021, biava2021, candini2023}). We note that, especially for bubbles D3 the maximum age presented may be underestimated because the oldest regions of the bubbles are too faint to be included in the age map and this implies that the jet activity times are underestimated as well.

It is very interesting to note that the maximum derived age for the plasma in bubble C1/C2, which is representative of the moment the jet started to inflate these bubbles, overlaps with the minimum derived age for the plasma in bubble D3, which is representative of the moment when particle acceleration within the lobe stopped. The same can be observed when comparing bubbles C1/C2 with bubbles B1/B2. As already suggested in \cite{brienza2021} purely based on morphological considerations, the continuity between the properties of the various generations of bubbles may imply either a scenario where the jets have actually switched off for a very short amount of time or, alternatively, a scenario in which new bubbles systematically detach from a continuously operating jet. The latter scenario can occur in case the jet power is low ($\rm Q_{jet}<5\times 10^{35}$ W, \citealp{luo2010}) so that the buoyancy velocity exceeds the jet-driven expansion even before the jet switch off. Discerning the two scenarios is observationally difficult but further discussion on this is presented in Sect. \ref{sec:expansion}.

We stress that, when interpreting the derived age values, one should remember that simple radiative ageing models describe particle populations that evolve passively after the initial injection with a constant and uniform magnetic field and without taking into account any possible adiabatic expansion of the lobes in the surrounding medium and any plasma reprocessing phenomena, such as re-acceleration. While this is possibly a reasonable enough assumption for many standard radio galaxies (e.g. \citealp{harwood2013}), in the case of Nest200047 it might be an oversimplification. Being located outside the IGrM core region, as estimated by X-ray observations \citep[see][]{brienza2021}, it is likely that bubbles C1/C2 and D3 suffer from non-negligible expansion losses. Moreover, the complex morphology of the radio emission implies that the bubbles are perturbed and the magnetic field is inhomogeneous. A more detailed discussion of this is presented in Sect. \ref{sec:expansion}.

In any case, the overall scenario described above is consistent with the idea that, in galaxy groups, the AGN jet activity must be gentle and quasi-continuous \citep[e.g.][]{gaspari2011, gaspari2014, bourne2021}. Owing to their shallow potential well, the impact of powerful AGN outbursts can be especially dramatic in these systems. The energy released can be significantly larger than what is required to offset IGrM cooling, causing a disruption of the core and a depletion of baryons within the virial radius \cite{ehlert2018}.

Despite this theoretical expectation, observing multiple generations of AGN bubbles in galaxy groups to probe the jet duty cycle has always been non-trivial, both in the radio band, where their non-thermal emission is observed, and in the X-ray band, where their imprints on the thermal IGrM is observed through cavities. Indeed, because this duty cycle takes place in a cosmological context, the signatures of multiple episodes of jet activity can be easily erased by fluid instabilities, mergers, and gas sloshing.

Moreover, for galaxy groups observations of such a phenomenon are even more difficult. Contrary to galaxy clusters where the dense ICM is likely able to maintain the non-thermal plasma confined and thus visible in the radio band for longer periods of time, in galaxy groups, the bubbles likely expand and disappear on much shorter timescales due to the lower IGrM density. At the same time, detecting cavities in galaxy groups in the X-ray band requires very long exposures with current instruments to achieve the necessary brightness contrast, especially outside the system's core region. The clearest case where this is observed is the galaxy group NGC\,5813, where within 20 kpc from the central source \cite{randall2011, randall2015} detected three pairs of collinear cavities likely produced by three distinct outbursts, suggesting a very quick duty cycle. Another famous example is the brightest X-ray group, NGC 5044, where deep X-ray observations have traced signs of (at least) three cycles of AGN outbursts; a pair of sub-kpc scale (inner) cavities close to the AGN, as well as older 5-kpc scale (intermediate) cavities \citep{Gastaldello2009,David2011,David2017}.  

Things have started to change now thanks to the advent of the new-generation radio telescopes that are sensitive to very old and low-surface brightness emission and this will further advance with the Square Kilometre Array Observatory (SKAO) and future generation X-ray observatories like ATHENA. What is certain is that, to date, Nest200047 remains a unique case where so many generations of outbursts can be probed directly in the radio band with such a high level of detail, possibly due to a combination of physical extent, line of sight, redshift, and duty cycle.

\subsection{Bubbles evolution}
\label{sec:expansion}

In this section, we explore whether the modelled spectral ages assuming only radiative cooling (see Sect. \ref{sec:time}) are consistent with the dynamical evolution of the bubbles, including their possible adiabatic expansion.
To assess the roles of adiabatic expansion and radiative losses we can make a consistency check of the observed spectral breaks with the expected values based on the model of buoyantly rising bubbles. 

\begin{figure}
\centering
\includegraphics[angle=0,trim=1cm 5.5cm 1cm 2.5cm,width=0.9\columnwidth]{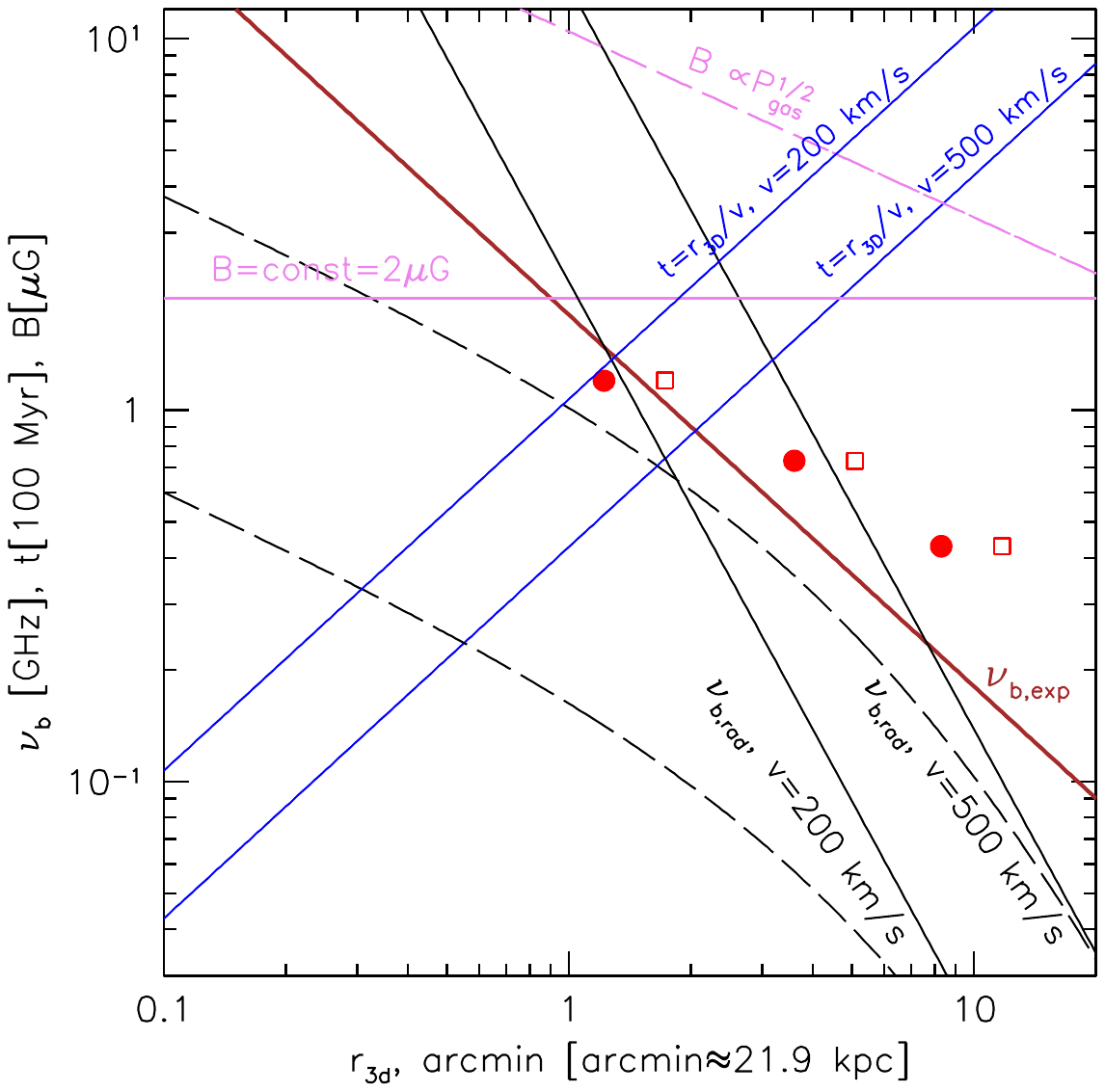}
\caption{Comparison of observed (red symbols) and expected break frequencies for regions B, C, and D3 as a function of their 3D radial distances ($r_{3D}$) from the cluster core. Two cases are shown: the solid circles and open squares correspond to $r_{3D}=r_{\rm proj}$ and $r_{3D}=r_{\rm proj}\times\sqrt{2}$, respectively. The brown curve shows the shift of the break frequency with distance due to pure adiabatic losses with $B\propto P_{\rm gas}^{1/2}$. The two blue lines show the characteristic dwell time (in units of $10^8\,{\rm yr}$) of a bubble at a given radius for two values of the bubble velocity. The black lines show the estimated break frequencies due to radiative losses for these velocities. The solid line is for constant $B$ and the dashed line is for $B\propto P_{\rm gas}^{1/2}$. The corresponding radial dependencies of $B$ are shown with the solid and dashed magenta lines for a constant and declining magnetic field, respectively.}
\label{fig:aging_bub}
\end{figure}

For this exercise, we used the observed radial distances and measured break frequencies of three regions "B", "C", and "D3" shown in Fig.~\ref{fig:aging_bub} with red symbols ($r_{\rm proj,B}$=1.1 arcmin, $r_{\rm proj,C}$=3.7 arcmin, and $r_{\rm proj,D3}$=8 arcmin). Since we do not know their true 3D positions, we show two cases: (a) all three regions are in the sky plane (solid circles) and (b) they are in the plane at 45 degrees to the line of sight (open squares, $r_{3D}=r_{\rm proj}\times\sqrt{2}$). 

To model the evolution of the radio spectra due to expansion, we use the thermal pressure profile derived from X-ray data (see \citealt{brienza2021} and \citealt{majumder2025}). As in \cite{churazov2001} we further assume that the energy density of the magnetic field inside the rising bubble makes a fixed fraction of the thermal pressure so that $B^2/8\pi$ plus other non-thermal components inside the bubble balances the outside thermal pressure. With these assumptions, $B(t)=B_0 C^{w}$, where $C=V(t=0)/V(t)$ is the inverse of the bubble expansion factor, $V(t)$ is the bubble volume, and $w=4/3$. If the bubble stays in pressure equilibrium with the ambient gas and its content can be described by a fluid with an adiabatic index of $4/3$, then $C^{4/3}\propto P_{\rm gas}$.

In the absence of acceleration mechanisms, the Lorentz factor changes as 
\begin{equation}
\gamma = \frac{\gamma_0 C^{1/3}}{1 + \gamma_0  b},
\label{eq:gamma_t}    
\end{equation}
where 
\begin{equation}
b = \frac{4 \sigma_T}{3mc} \int_{t_0}^{t} \left (\frac{B_0^2}{8 \pi}C^{4/3}(t') + u_{IC}\right )C^{1/3}(t')d t',
\label{eq:b}    
\end{equation}
where $u_{IC}$ is the energy density of the radiation field (= CMB), \citep[e.g.][]{kardashev1962}. Then, the break in the electron's distribution  is at 
$\gamma_b\sim \frac{C^{1/3}}{b}$ and the corresponding break in the observed spectrum is at $\nu_b\propto \frac{C^{4/3}}{b^2}$. 
Pure adiabatic expansion does not introduce itself any break in the distribution function of relativistic electrons but merely shifts the existing break, if any, in the observed spectrum due to a change of $B$ and $\gamma_b$ associated with the expansion factor $C$.

In Fig.~\ref{fig:aging_bub}, we show different model lines describing the evolution of various quantities as a function of distance from the group centre: break frequency $\nu_b$, age $t$, and magnetic field $B$. The blue lines represent the bubble's age $t=r_{3D}/v$ as a function of radius for two different speed values: $v$=200~km~$\rm s^{-1}$ and $v$=500~km~$\rm s^{-1}$. The latter value yields the buoyancy time derived for bubble D3 in \cite{brienza2021} and is lower than the sound speed in the IGrM $c_{s}$=720 km $\rm s^{-1}$. The black lines show the expected break frequency for these two values of the velocity, assuming pure radiative ageing in a constant magnetic field (solid line) and a magnetic field declining with radius (dashed lines). For the latter case, $B$ is assumed to be equal to 9 $\mu$G at the position of bubble B, which can be considered a robust upper limit inferred from the gas pressure in the group core.
Finally, the solid brown line represents the shift of the break frequency with distance due to pure adiabatic losses with $B\propto P_{\rm gas}^{1/2}$, normalised to the initial break frequency observed in the bubble B, $\rm\nu_b\sim 1.2\,{\rm GHz}$. Based on this, a bubble moving from the core to positions of regions C and D3 will have the breaks at $\sim 0.6$ and $\sim 0.3$ GHz only due to expansion.

From the plot, we can infer that the break frequency observed in bubble B does not contradict pure radiative age models with either $B$=const and $v$=200 km $\rm s^{-1}$ or decaying $B$ with $v$=500 km $\rm s^{-1}$. What is observed for bubble C is marginally consistent with either a pure expansion model or with pure radiative ageing for $B$=const or decaying $B$ and $v$=500 km $\rm s^{-1}$. Instead, D3 appears in tension with all models, even in the minimal distance case (when the bubble moves in the sky plane) and especially if we consider that the contributions of adiabatic expansion and radiative ageing should in reality be combined and this leads to a further steepening of the spectrum (by at least a factor of a few with respect to the pure ageing scenario).

This exercise suggests that it is hard to consider bubbles B, C, and D as an evolutionary sequence of a single bubble. But if they are, the expected losses for the most external bubble D3 in this model are too large. In this case, radiative and expansion losses would need to be balanced by an extra mechanism, such as particle re-acceleration, unless expansion is slowed down by mixing of the bubble with the surrounding gas. Alternatively, the different bubbles, may be the result of subsequent jets outbursts with different properties.

\subsection{Filaments}
\label{sec:filam}

One of the most striking observed morphological features in Nest200047 is surely the complex array of non-thermal filamentary structures. In \cite{brienza2021} these were mainly identified in the outer regions of the system and interpreted as the result of the oldest lobe breaking apart while buoyantly rising in the group atmosphere and as a result of a turbulent environment. With the new high-resolution observations, we detect additional filaments in the inner $\sim$60 kpc as well, surrounding and departing from the inner bubbles and jets (structures A and B1/B2), suggesting a complex dynamical evolution for the younger lobes located in the group core as well. 

Similar elongated structures have been detected in various systems with sensitive, high-resolution radio observations. They have been found in different astrophysical environments such as the Galactic centre \citep{yusefzadeh2004, heywood2022b, yusefzadeh2022a, yusefzadeh2022b}, within radio galaxy lobes \citep[e.g.][]{fomalont1989, maccagni2020, mckinley2022, mahatma2023} and tails \citep[e.g.][]{gendronmarsolais2020, botteon2020, koribalski2024}, departing from radio galaxy lobes/jets \citep[e.g.][]{ramatsoku2020, rudnick2022, condon2021, velovic2023} within relics in galaxy clusters \citep[e.g.][]{owen2014, degasperin2022, rajpurohit2023, churazov2023} and within radio phoenixes \citep[e.g.][]{mandal2020, pasini2022, raja2024}. However, detailed analysis of such structures are still limited and physical scenarios for their origin still under discussion \citep[see e.g.][] {yusefzadeh2022c, ruszkowski2023, churazov2023, ohmura2023}.

Understanding the 3D morphology of these filaments remains challenging. Based on cosmological simulations of radio relics \cite{wittor2023} suggests that the filaments observed in these sources are not produced by sheets seen in projection, but rather by actual filaments and ribbons. In the case of Nest200047 this may possibly be also the case, at least for the majority of the filaments (F1-F7). However, for filament D2, one possibility is that it represents a ring as seen in projection, as predicted by models of bubble evolution. In this occurrence, the two sub-filaments observed on the Eastern side of D2 could be interpreted as the front and back of such ring viewed in projection. Moreover, the boxy filament, whose morphology is among the most puzzling, may actually result from the projection of a 3D layer enveloping the entire bubble D3. Sharp edges in the radio emission of radio lobes can indeed form due to magnetic draping as demonstrated by numerical simulations and observations of AGN bubbles \citep{ruszkowski2008, adebahr2019}. Polarisation observations could provide further insight into this.

The filaments observed in Nest200047 show a large variety of lengths, ranging from a few tens of kpc up to a few hundreds of kpc, while they mostly appear unresolved or barely resolved along their width, implying sizes of a few kpc and down to 2 kpc. The only filament that clearly appears spatially resolved along its width (with size up to $\sim$18 kpc) is D2, which, however, based on our high-resolution imaging, might be, in reality, the result of the superposition of two thinner and very close separate components, as especially visible on the Eastern side. 

These ranges of dimensions are similar to those of other filaments reported in the literature showing a variety of lengths (e.g. 220 kpc, \citealp{rudnick2022}; 30-50 kpc, \citealp{condon2021}; $\sim$60 kpc, \citealp{brienza2022}; $\sim$40 kpc, \citealp{mahatma2023}; 20-50 kpc, \citealp{candini2023}; $\sim$160 kpc, \citealp{velovic2023}; $\sim$200 kpc \citealp{botteon2024}) and approximately a similar width equal to a few kpc, again possibly limited by the spatial resolution of the observations.

As discussed in \cite{brienza2021} the Alfvén scale in this system is of the order of a few kpc, consistent with the average width of filaments. This suggests that below these scales magnetic fields start to play a key role in the dynamical evolution of the filaments, and possibly prevent them from being destroyed by hydrodynamic instabilities. Moreover, \cite{rudnick2022} points out that the thickness of a few kpc observed for these filaments is likely to represent the size of bundles of smaller “fibres”, which are thought to arise naturally in turbulent magnetic plasmas \citep{porter2015}.

It is also interesting to note that these filaments often come in pairs or groups \citep{rudnick2022, botteon2024, yusefzadeh2022a, condon2021}. In the case of the Galactic filaments \cite{yusefzadeh2022a} speculates that the filamentation is the result of an interaction with an obstacle, which sets the length scale of the separation between the filaments. In Nest200047 we see two pairs of filaments on large scales F1/F2 and F4/F7 both with a separation of $\sim$25 kpc and one pair in the central region F5/F6, which seem to intersect at the bottom, possibly in projection.

As far as the spectral properties are concerned, our analysis shows that all filaments exhibit steep and curved spectra with $\rm \alpha>1$, consistent with being old plasma. It is interesting to note, however, that they can have different spectral index values even if located very close, possibly indicating variations in magnetic field strengths rather than in ages (see e.g., the difference between F5 and F6, best highlighted in the tomography maps in Fig \ref{fig:tomo1}). 

As shown in Sect. \ref{sec:bright}, within each filament, regions with higher surface brightness also tend to have flatter spectral indices and more pronounced curvature. This behaviour is similar to what is observed in radio relics \citep{rajpurohit2022}, where the brightest regions show the flattest spectral index, interpreted as traces of shocks with the highest Mach numbers and thus the sites of particle acceleration. However, in the case of relics, no significant spectral curvature is reported in those regions. We also notice that the strength of the correlation between brightness and spectral index varies significantly with the observing frequency, becoming more and more pronounced moving to higher frequencies.

One notable result is that for most filaments the spectral index does not show any strong gradient along their length, instead it appears to be uniform with a typical variation of $\rm \Delta\alpha\sim0.2$. The only filament where we possibly see a trend is D2, where we observe flatter values in the Eastern regions with respect to the Western ones, but still with maximum $\rm \Delta\alpha\sim0.3$. However, we remind the reader again, that these structure might be produced by a superposition of narrower components that might affect this conclusion.

A similar flat spectral trend is observed in filaments in other sources as well, such as \cite{brienza2022, giacintucci2022, candini2023, digennaro2024}. However, other cases do also exist where a significant and near-monotonic spectral steepening with distance along the filament is observed such as \cite{rudnick2022} (from $\rm \alpha\sim1$ to $\rm \alpha\sim3$ over 220 kpc), \cite{velovic2023} (from $\rm \alpha\sim2$ to $\rm \alpha\sim4$ over $\sim$160 kpc) and \cite{raja2024} (from $\rm \alpha\sim1.7$ to $\rm \alpha\sim2.6$ over $\sim$170 kpc). Such a steepening could be consistent with filaments representing open flux tubes through which cosmic rays escape from the main radio galaxy body and age, but this is not observed in filaments F1-F6 in Nest200047, which may enter this category.

The absence of a spectral index gradient along the filaments, up to hundreds of kpc is not trivial to explain. In a scenario in which these filamentary structures are the result of AGN bubble fragmentation during its rise in the group atmosphere, possibly combined with the system's weather, as proposed in \cite{brienza2021}, the easiest interpretation would probably be that particles within the filaments have been accelerated approximately within a single event and evolve in a similar magnetic field, so they show similar spectral shapes.

Alternatively, particles along the filaments could be (re-)accelerated. The electrons spiralling in amplifying magnetic fields with varying magnetic field are subjected to an electromotive force, allowing them to be (re-)accelerated, for instance, via the Betatron mechanism (see \citealp{melrose1980}, for a review). In the simplest applicable scenario where the magnetic field increases on a time scale shorter than the cooling time of the electrons, and while the isotropy of the electron pitch angle is maintained through scattering, the energy increment of the electrons can be expressed as 

\begin{equation}
(\Delta p_c)^2 \sim \frac{2}{3} (p_c)^2 \frac{\delta B}{B} = \frac{2}{3} (p_c)^2 \phi,
\end{equation}
where $\phi = \frac{\delta B}{B}$.

The simultaneous increase in both the magnetic field strength and electron energy results in a relative increase in the characteristic synchrotron emission frequency.

We note that a significant challenge to a scenario in which particles move along the filaments to reproduce the observed uniform spectral trend is the necessity for diffusion to occur at a sufficiently rapid rate. Specifically, the electron scattering rates must be slow enough so that electrons can propagate at least 100 kpc along filaments in a time-scale that is shorter than or comparable to their radiative cooling time, $\tau_{\text{rad}} \sim 50 \, \text{Myr}$. This condition would imply strongly super-Alfvenic streaming along the filaments, establishing a lower limit on the effective scattering time-scale  $\tau_s \sim 10^4 \, \text{yrs}$.

\section{Summary and conclusions}
\label{sec:concl}

By combining data from our large observational campaign in the frequency range 53-1518 MHz, including LOFAR, uGMRT, MeerKAT and VLA observations, we performed a detailed morphological and spectral analysis of the galaxy group Nest200047, which provided an unprecedented view of this intriguing system. 
Our main findings are as follows:

\begin{itemize}

\item The overall morphology of the radio emission previously observed in the system at frequencies $<$144 MHz is recovered up to 1518 MHz, except for the very large-scale steep-spectrum emission. New filamentary structures are revealed within the inner $\sim$60 kpc of the system, surrounding and departing from the inner bubbles and jets (structures A and B1/B2), suggesting a complex dynamical evolution also for the younger lobes located in the group core. Overall, we detect a lot of filamentary emission across all scales, with lengths ranging from a few kpc up to 350 kpc and width always of few kpc.

\medskip
    
\item At high ($\sim$5 arcsec) resolution, the main filament D2, located in the northern outer bubble, appears to be composed of two substructures, especially on the Eastern side. This may indicate the presence of two closely aligned 2D filaments that are partially overlapping. Alternatively, a more intriguing interpretation is that they represent the back and front of a 3D ring seen in projection, as predicted by models of bubble evolution.

\item All filaments are characterized by steep and curved spectra with $\rm \alpha>1$, consistent with being old plasma. We do not measure a significant spectral index trend along the filaments. This might imply that particles in the filaments were accelerated approximately within a single event and evolved in a similar magnetic field or were re-accelerated. Any scenarios involving particles movement along the filaments would require super-Alfvenic speed streaming to produce the observed flat spectral profile.

\medskip

\item By performing spectral ageing analysis with a JP model and assuming the magnetic field value providing minimum radiative losses equal to 1.95 $\rm \mu$G, we derive that the median radiative ages of the plasma in the three generations of bubbles B1/B2, C1/C2, and D3 are 130 Myr, 160-170 Myr, and >220 Myr, respectively. A first-order estimate of the duration of particle acceleration within each bubble suggest $t$>50 Myr. The continuity in the spectral properties across the various generations of bubbles may imply either a scenario in which the jets switched off for only a very short time or, alternatively, a scenario in which new bubbles systematically detach from a continuously operating jet. In either case, the observations seem to be consistent with the assumption of a quasi-continuous energy injection as suggested in the "maintenance" mode of the AGN feedback.

\medskip

\item The spectral analysis, performed using a combination of colour-colour plots, global spectra and spectral age mapping, reveals that aligning the spectra of regions from different bubbles and the main filament D2 to a single spectral curve is challenging (i.e. the injection index appears different). This implies differences in the acceleration processes across bubbles and/or variations in magnetic field fluctuations or particle reprocessing.

\medskip

\item The energy losses expected for the outermost northern bubble D3, including both radiative ageing and adiabatic expansion, and assuming a dynamical evolution from bubble B1/B2 appear to be too large with respect to what inferred observationally from the spectral break. This suggests that if bubbles B, C, and D are considered an evolutionary sequence of a single bubble, re-acceleration of particles in the outermost bubble D3 has to be invoked to balance the expected radiative and expansion losses, unless expansion is slowed down by mixing of the bubble with the surrounding gas. Alternatively, the different bubbles, may be the result of subsequent jets outbursts with different properties.

\end{itemize}

In conclusion, our detailed broad-band spectral analysis allowed us to make new steps forward in the comprehension of Nest200047, which represents an incredible laboratory to study recurrent jet activity and feedback in a galaxy group, and, more in general, highlight the big potential of the combined use of data from SKA precursors and pathfinders. 

Many aspects of this complex system remain a big puzzle, which challenges our understanding, and will require further observational investigations and simulations. For example, \textit{'How exactly is the bizarre boxy shape filament formed and maintained, as well as all other filaments? What makes this group special with respect to the many others present in the sky?} At present, the LOFAR Two-meter Sky Survey (LoTSS; \citealp{shimwell2019, shimwell2022}) covers $\sim$85\% of the Northern sky, and no other radio galaxy has been found to share a similar morphology. Possibly, a fortunate combination of redshift, line of sight, duty cycle and ambient medium is what drives its uniqueness. 

Some of these aspects will be addressed in upcoming papers focussing on the analysis of the X-ray thermal emission \citep{majumder2025} and the radio polarisation properties (Brienza et al. in prep.).

\begin{acknowledgements}

MBrienza acknowledges financial support from Next Generation EU funds within the National Recovery and Resilience Plan (PNRR), Mission 4 - Education and Research, Component 2 - From Research to Business (M4C2), Investment Line 3.1 - Strengthening and creation of Research Infrastructures, Project IR0000034 – “STILES - Strengthening the Italian Leadership in ELT and SKA”. MBrienza acknowledges financial support from the Italian L’Oreal UNESCO "For Women in Science" program and from the ERC-Stg “MAGCOW", no. 714196, financial support from the agreement ASI-INAF n. 2017-14-H.O, from the PRIN MIUR 2017PH3WAT "Blackout" and from the INAF RF 2023. 
KR acknowledges the Smithsonian Combined Support for Life on a Sustainable Planet, Science, and Research administered by the Office of the Under Secretary for Science and Research.
IKhabibullin acknowledges support by the COMPLEX project from the European Research Council (ERC) under the European Union’s Horizon 2020 research and innovation program grant agreement ERC-2019-AdG 882679.
MBr\"uggen acknowledges funding by the Deutsche Forschungsgemeinschaft (DFG) under Germany's Excellence Strategy -- EXC 2121 ``Quantum Universe" --  390833306 and the DFG Research Group "Relativistic Jets".
FV acknowledges Fondazione Cariplo and Fondazione CDP, for grant n° Rif: 2022-2088 CUP J33C22004310003 for "BREAKTHRU" project.

The MeerKAT telescope is operated by the South African
Radio Astronomy Observatory, which is a facility of the National Research Foundation, an agency of the Department of Science and Innovation. We wish to acknowledge the assistance of the MeerKAT science operations team in both preparing for and executing the observations that have made our census possible.

We acknowledge the use of the ilifu cloud computing facility - \url{www.ilifu.ac.za}, a partnership between the University of Cape Town, the University of the Western Cape, Stellenbosch University, Sol Plaatje University, the Cape Peninsula University of Technology and the South African Radio Astronomy Observatory. The ilifu facility is supported by contributions from the Inter-University Institute for Data Intensive Astronomy (IDIA - a partnership between the University of Cape Town, the University of Pretoria and the University of the Western Cape), the Computational Biology division at UCT and the Data Intensive Research Initiative of South Africa (DIRISA).

LOFAR, the Low Frequency Array designed and constructed by ASTRON, has facilities in several countries, which are owned by various parties (each with their own funding sources), and are collectively operated by the International LOFAR Telescope (ILT) foundation under a joint scientific policy. The ILT resources have benefited from the following recent major funding sources: CNRS-INSU, Observatoire de
Paris and Universit\'e d'Orl\'eans, France; BMBF, MIWF-NRW, MPG, Germany;
Science Foundation Ireland (SFI), Department of Business, Enterprise
and Innovation (DBEI), Ireland; NWO, The Netherlands; the Science and
Technology Facilities Council, UK; Ministry of Science and Higher
Education, Poland.

Part of this work was carried out on the Dutch national
e-infrastructure with the support of the SURF Cooperative through
grant e-infra 160022 \& 160152. The LOFAR software and dedicated
reduction packages on \url{https://github.com/apmechev/GRID_LRT} were
deployed on the e-infrastructure by the LOFAR e-infragroup, consisting
of J.\ B.\ R.\ Oonk (ASTRON \& Leiden Observatory), A.\ P.\ Mechev (Leiden
Observatory) and T. Shimwell (ASTRON) with support from N.\ Danezi
(SURFsara) and C.\ Schrijvers (SURFsara). This research has made use of the University
of Hertfordshire high-performance computing facility
(\url{https://uhhpc.herts.ac.uk/}) and the LOFAR-UK compute facility,
located at the University of Hertfordshire and supported by STFC
[ST/P000096/1]. This research made use of the LOFAR-IT computing infrastructure supported and operated by INAF, and by the Physics Dept. of Turin University (under the agreement with Consorzio Interuniversitario per la Fisica Spaziale) at the C3S Supercomputing Centre, Italy. 

The J\"ulich LOFAR Long Term Archive and the German
LOFAR network are both coordinated and operated by the J\"ulich
Supercomputing Centre (JSC), and computing resources on the
supercomputer JUWELS at JSC were provided by the Gauss Centre for
supercomputing e.V. (grant CHTB00) through the John von Neumann
Institute for Computing (NIC). 

We thank the staff of the GMRT that
made these observations possible. GMRT is run by the National Centre for Radio Astrophysics of the Tata Institute of
Fundamental Research.

The National Radio Astronomy Observatory is a facility of the National Science Foundation operated under cooperative agreement by Associated Universities, Inc.

This research made use of astropy, a community-developed core Python package for astronomy \citep{astropy2013, astropy2018, astropy2022} hosted at \url{http://www.astropy.org/}.

\end{acknowledgements}

\bibliographystyle{aa}
\bibliography{nest-paper2.bib}

\begin{appendix}
\onecolumn
\section{Spectral index uncertainty maps}
\label{appendix1}

\begin{figure*}[!htp]
\centering
\includegraphics[width=0.6\textwidth]{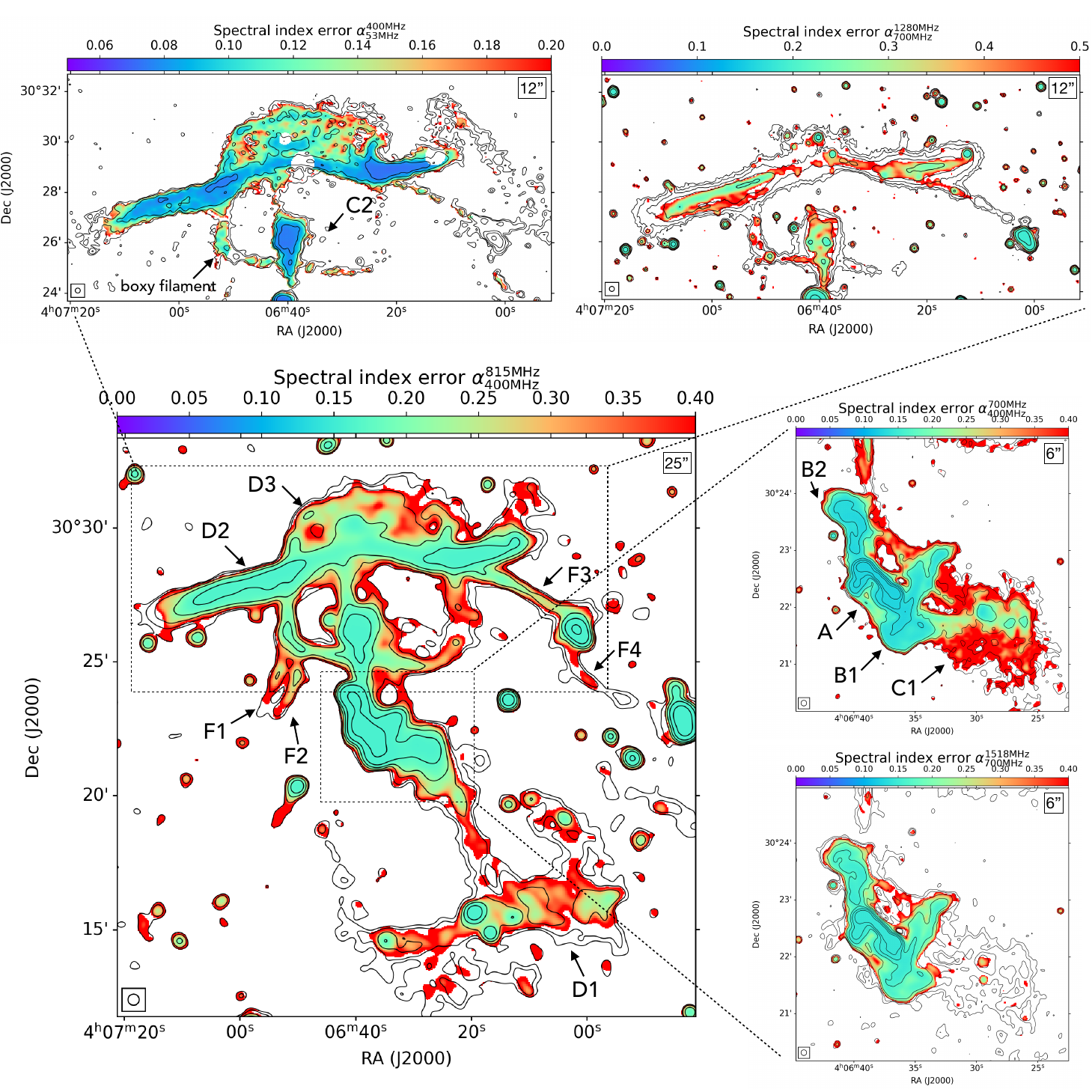}
\caption{Spectral index uncertainty maps of Nest200047. \textit{Central panel}: spectral index uncertainty map of the entire source between 400-815 MHz with a resolution of 25 arcsec. \textit{Top panels}: spectral index uncertainty maps of the northern bubbles between frequencies 53-400 MHz (left), and 700-1280 MHz (right), with a resolution of 12 arcsec. \textit{Right panels}: spectral index uncertainty maps of the central jets/bubbles between 400-700 MHz (top) and 700-1518 MHz (bottom) with a resolution of 6 arcsec. Black contours always trace the emission at the lowest frequency included in the map starting from $\rm 3\sigma$. See Sect. \ref{sec:spix} for a full description of the derivation.}
\label{fig:shift}
\end{figure*}

\section{Spectral age fits}
\label{appendix2}

\begin{figure*}[!htp]
\centering
\includegraphics[width=0.8\textwidth]{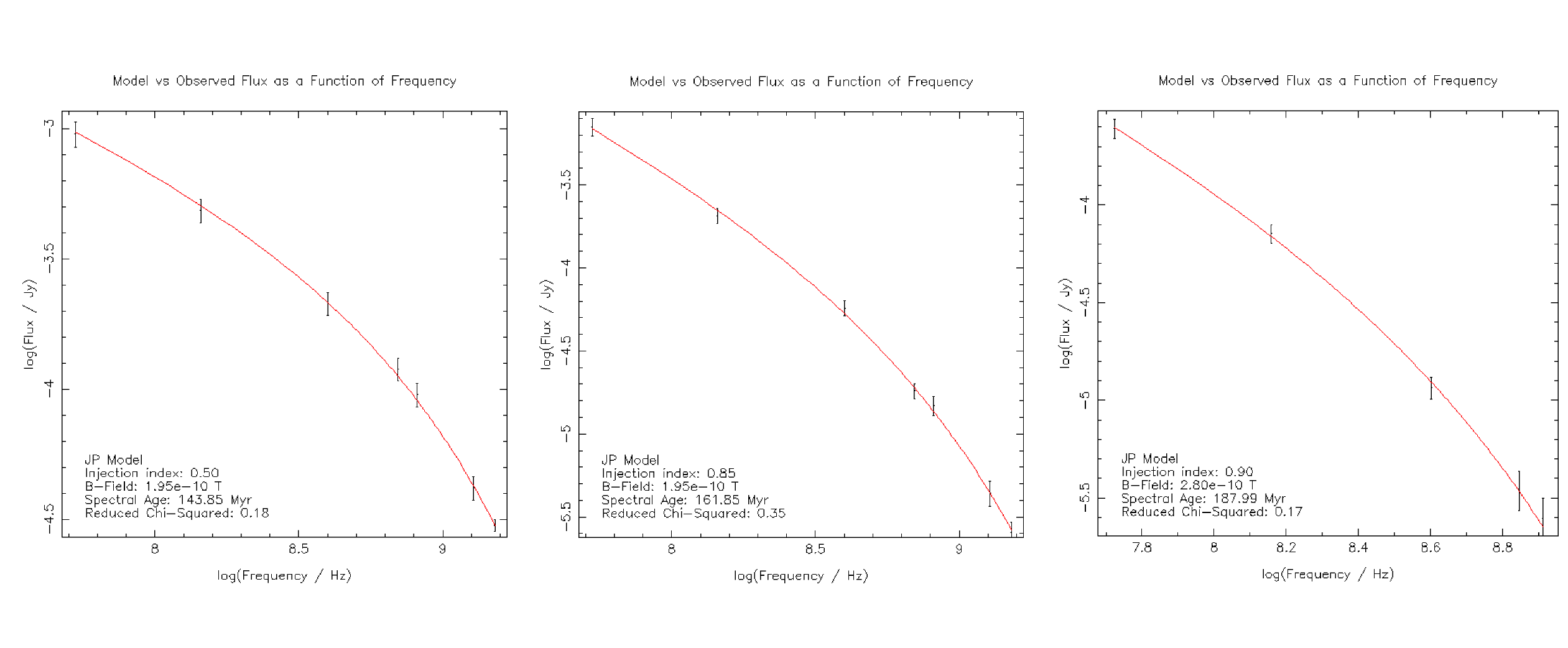}
\caption{Example of spectral fits for three individual pixels belonging to the three different regions analysed: B1/B2 (left), C1/C2 (middle) and D3 (right). See Sect. \ref{sec:age} for a full description of the modelling approach.}
\label{fig:shift}
\end{figure*}
\end{appendix}

\end{document}